\documentclass[usenatbib,twocolumn]{mnras}

\usepackage[T1]{fontenc}
\usepackage{ae,aecompl,color,footnote}
\usepackage{graphicx}      
                               
\usepackage{epstopdf}
\usepackage{amsmath}    
\usepackage{amssymb}    
\usepackage{booktabs}
\usepackage{bm}
\usepackage{tabularx}                                                                                                              
\usepackage{multicol}                                                                                                              
\usepackage[normalem]{ulem}

\usepackage{float}
\def\apj{ApJ}%
\def\apjl{ApJ Lett.}%
\def\aap{A\&A}%
\def\mnras{MNRAS}%
\def\physrep{Phys. Rep.}%
\def\prc{Phys. Rev. C}%
\def\prd{Phys. Rev. D}%
\def\nat{Nature}%
\def\nphysa{Nucl. Phys. A}%
\def\physrep{Phys. Rep.}%
\usepackage[normalem]{ulem} 

\title[Scaling in hot stellar matter]
      {Proto-neutron stars with heavy baryons and
        universal relations}

      \author[A. R. Raduta et al.]{
        Adriana R. Raduta$^{1}$\thanks{araduta@nipne.ro},
        Micaela Oertel$^{2}$\thanks{micaela.oertel@obspm.fr}
        and Armen Sedrakian$^{3,4}$\thanks{sedrakian@fias.uni-frankfurt.de} 
\\
$^{1}$National Institute for Physics and Nuclear Engineering
(IFIN-HH),  RO-077125 Bucharest, Romania\\
$^{2}$LUTH, Observatoire de Paris, Universit\'e PSL,
CNRS, Universit\'e de Paris, 92195 Meudon, France\\
$^{3}$Frankfurt Institute for Advanced Studies, D-60438 
Frankfurt-Main, Germany  \\
$^{4}$Institute of Theoretical Physics,  University of Wroc\l{}aw,
50-204 Wroc\l{}aw, Poland
}

\begin{document}
\label{firstpage}
\pagerange{\pageref{firstpage}--\pageref{lastpage}}
\maketitle

\begin{abstract}
  We use covariant density functional theory to obtain the equation of
  state (EoS) of matter in compact stars at non-zero temperature,
  including the full baryon octet as well as the $\Delta(1232)$
  resonance states. Global properties of hot $\Delta$-admixed
  hypernuclear stars are computed for fixed values of entropy per baryon ($S/A$) and lepton fraction ($Y_L$). Universal relations between the moment of inertia, quadrupole moment, tidal deformability, and compactness of compact stars are established for fixed values of $S/A$ and $Y_L$ that are analogous to those known for cold catalyzed compact stars. We also verify that the
  $I$-Love-$Q$ relations hold at finite temperature for constant values of $S/A$ and $Y_L$.
\end{abstract}

\maketitle

\section{Introduction}
\label{sec:intro}

Cold, mature compact stars are well described by a one-parameter
equation of state (EoS) relating pressure to (energy) density.  In
contrast, the studies of the dynamics of core-collapse supernovae
(CCSN)
\citep{Janka_PhysRep_2007,Mezzacappa2015,Connor2018ApJ,Burrows2020MNRAS},
proto-neutron star (PNS) evolution \citep{Pons_ApJ_1999}, stellar
black-hole (BH) formation
\citep{Sumiyoshi_2007,Fischer_2009,OConnor_2011,hempel12} and binary
neutron star (BNS) mergers
\citep{Shibata_11,Rosswog_15,Baiotti_2017,Ruiz2020} require as an
input an EoS at non-zero temperature and out of (weak)
$\beta$-equilibrium, {\it i.e.}, the pressure becomes a function of
three thermodynamic parameters. For describing all of the above
mentioned astrophysical scenarios one needs to
consider baryon number densities, $n_B$, ranging from sub-saturation
densities up to several times the nuclear saturation density
$n_s \simeq 0.16$~fm$^{-3}$, temperatures up to 100 MeV and charge
fractions $0 \leq Y_Q = n_Q/n_B \leq 0.6$, where $n_Q$ is defined as
the total hadronic charge density~\citep{Oertel_RMP_2017}.

In recent years the EoS of cold compact stars has been considerably
constraint due to several new astrophysical observations (compact star
masses, radii and tidal deformability), information coming from
experimental nuclear physics and the progress in {\it ab-initio}
calculations of pure neutron matter.  The new data has, in particular,
narrowed the possible parameter space of EoS derived from density
functional theory (DFT). There exist a large number of DFT based EoS
which are applicable to cold $\beta$-equilibrated neutron stars, see
{\it e.g.}
\citep{Weissenborn2012a,Weissenborn2012b,Colucci_PRC_2013,Dalen2014,Oertel2015,Chatterjee2015,Fortin_PRC_2016,Chen_PRC_2007,Drago_PRC_2014,Cai_PRC_2015,Zhu_PRC_2016,Sahoo_PRC_2018,Kolomeitsev_NPA_2017,Li_PLB_2018,
  Li2019ApJ,Ribes_2019,Li2020PhRvD}.  The latter works address, in
particular, the problem of hyperonization of dense matter ({\it the
  hyperon puzzle}) and the possible emergence of $\Delta$-degrees of
freedom in dense matter at zero temperature.  Models of finite
temperature dense matter with heavy baryon degrees of freedom,
applicable to CCSN and BNS mergers, which are consistent with the
constraints imposed by modern data are less numerous, see
\citep{Oertel_PRC_2012,Colucci_PRC_2013,Oertel_EPJA_2016,Marques_PRC_2017,Dexheimer_2017,Fortin_PASA_2018,Malfatti_PRC_2019,Stone_2019}.

The first aim of this work is to obtain an EoS of non-zero temperature
matter, which is well-constrained by astrophysics and laboratory data,
on the basis of the DDME2 parameterisation~\citep{Lalazissis_PRC_2005}
extended to hypernuclear matter by
\cite{Fortin_PRC_2016}\footnote{Similar extensions of the DDME2
  parametrisation to the hyperonic sector, which differs in the way
  the parameters in the scalar-meson sector are fixed, can be found in
  \cite{Colucci_PRC_2013,Dalen2014,Li_PLB_2018}.}.  We include
additionally the $\Delta$ degrees of freedom as they add an important
feature: they soften the EoS in an intermediate density range, reduce
the compact star radii for intermediate
masses~\citep{Drago_PRC_2014,Li_PLB_2018} which improves the agreement
between theoretical models and observations. The resulting
finite-temperature EoS are relevant for the variety of above mentioned
astrophysical scenarios, which are known to depend sensitively on the
EoS
\citep{Pons_ApJ_1999,Sekiguchi_PRL_2011,Bauswein_2012,Fischer2014,Perego_EPJA_2019,Peres_2014,Schneider_2019,Bauswein_2019,Schneider_2020,Yasin_2020}.

To illustrate the properties of the EoS and the impact of heavy
baryons, we make calculations for constant values of entropy per
baryon $S/A$ ($0 \leq S/A \leq 4$) either assuming constant lepton
fraction $Y_L$ ($0.2 \leq Y_L \leq 0.4$) in the neutrino-trapped
regime or freely streaming neutrinos and matter in
$\beta$-equilibrium.  Some of the considered thermodynamic conditions
are relevant for different stages in the evolution from a PNS to a
compact star \citep{Burrows_ApJ_1986,Prakash_1997,Pons_ApJ_1999}.
Broad coverage of all the astrophysics relevant parameter space will
be given elsewhere. Under the conditions of PNS, the neutrinos are
trapped and the lepton fraction is thus given by the sum of charged
leptons and neutrinos fractions. For each set of thermodynamic
conditions, we will discuss properties of the EoS and matter
composition as well as maximum mass and radii of hot stars.  Possible
instability windows of non-accreting PNS with respect to collapse to a
BH is investigated following \cite{Bombaci_AA_1996}.  The maximum
gravitational mass at constant $S/A$, assuming neutrino-transparent
$\beta$-equilibrated matter, is related to the collapse to a BH during
in failed CCSNe \citep{Schneider_2020}. We study here the influence of
heavy baryons on this maximum mass.

The second motivation of this study is to test the universal relations
among the global properties of stationary, slowly and rigidly rotating
compact stars at finite temperature.  In essence, universality means
that these relations, well established at zero temperature, are {\it
  independent of the underlying
  EoS}~\citep{Yagi_PRD_2013,Maselli_PRD_2013,Breu_MNRAS_2016,Yagi2017PhR,Paschalidis_PRD_2018}
and are thus very helpful for the interpretation of observational
data since they allow to mitigate the uncertainties related to the
EoS.  Given the importance of such relations, it is interesting to
investigate them under new conditions, such as finite temperatures and
out-of-equilibrium with respect to weak interactions.

The possibility of some of these relations being universal at finite
temperature has been studied
before~\citep{Martinon_PRD_2014,Marques_PRC_2017,Lenka_JPG_2019}.  The
moment of inertia as a function of compactness was studied by
\cite{Lenka_JPG_2019}, who concluded that thermal effects lead to
deviations from the universal relations obtained for
$\beta$-equilibrated matter at zero temperature.  Similar conclusions
have been reached for the so-called $I-$Love$-Q$ relations, where
again it was claimed that at finite temperature deviations from the
zero-temperature universality are
found~\citep{Martinon_PRD_2014,Marques_PRC_2017}.  Below we show that,
if one considers fixed values of $(S/A,Y_{L,e})$, where  $Y_{L,e}$ is
the electron fraction, the
$\bar I-\bar \lambda$ and $\bar I-\bar Q$ relations are universal to
accuracy comparable to that obtained for cold compact stars. The
same is true for several global properties of compact stars, such as
the moment of inertia, quadrupole moment, tidal deformability as
function of compactness. Only the binding energy per unit of
the gravitational mass shows deviations from universality.

This paper is organized as follows.  In Sec.~\ref{sec:model} we
describe the details of the parameterisation that has been used in
modeling the EoS from relativistic density functional theory.
Section~\ref{sec:NS} is devoted to the discussion of the results for
the mass and radius of hot compact stars.  In
Section~\ref{sec:scalingC} we investigate the behavior of different
(normalised) global properties as a function of the star's
compactness.  Section~\ref{sec:scaling} focuses on $I-$Love$-Q$
relations.  We conclude in Sec.~\ref{sec:conclusions}. Throughout this
paper we use the natural units with $c=\hbar=k_B=G=1$.

\section{The model}
\label{sec:model}

We consider matter at non-zero temperature as it occurs
in CCSN, PNS and BNS mergers. Leptons  (electrons, muons and neutrinos)
and photons  are considered as non-interacting
gases,  whereas under the relevant conditions, the
tau lepton can safely be neglected. The partition function of the
system thus factorizes  into a product of baryonic
(index $B$), leptonic (index $L$) and photonic ($\gamma$) partition
functions: ${\cal Z} = {\cal Z}_B {\cal Z}_L {\cal Z}_{\gamma}$. In
thermodynamic equilibrium, within the
grand-canonical ensemble, 
\begin{eqnarray}
\label{eq:Z_leptons}
\frac{1}{V}\ln {\cal Z}_L &=& 
\sum_{j} \! g_j
\int\! \frac{d^3k}{(2 \pi)^3} \frac{k^2}{\varepsilon_j(k)} \nonumber \\
  && \left[ f_{\mathrm{FD}}(\varepsilon_j(k)-\mu_j) +  f_{\mathrm{FD}} (\varepsilon_j(k)+\mu_j) \right],\\
  \frac{1}{V}\ln {\cal Z}_{\gamma} &=&g_\gamma
  \int \frac{d^3k}{(2 \pi)^3} \frac{k^2}{\epsilon_{\gamma}(k)} f_{\mathrm{BE}}(\epsilon_{\gamma}(k)). 
\label{eq:Z_photons}  
\end{eqnarray}
{where} $f_{\mathrm{FD}}$ and $f_{\mathrm{BE}}$ {are} the Fermi-Dirac
and Bose-Einstein distribution functions at temperature $T$,
respectively, $g_j=(2s_j+1)$ is the spin degeneracy factor with $s_j$
{being} the lepton's spin [$g_j=2$ for charged leptons and $g_j=1$ for
(left-handed) neutrinos] {and} $\varepsilon_j(k)$ and $\mu_j$ {are} the
single particle energy and the chemical potential of the lepton $j$.
In Eq.~\eqref{eq:Z_photons} $g_{\gamma} = 2$ and
$\epsilon_{\gamma}(k)$ are the photon spin degeneracy and energy.  The
partition function of the strongly interacting baryons requires a
model of the strong nuclear interaction.  Within the relativistic
density functional theory it is given by a sum of kinetic part (which
has a form analogous to that  of an ideal gas) and a potential
term. We write it as
\begin{eqnarray}
\label{eq:Z_baryons}
\frac{1}{V}\ln {\cal Z}_B &=&\frac{1}{V }\ln~{\cal Z}_m
+\sum_{i\in \mathit{baryons}} \!\frac{2J_i+1}{3}
  \int\!\! \frac{d^3k}{(2 \pi)^3} \frac{k^2}{E_i(k)}  \nonumber \\
&\cdot& \left[ f_{\mathrm{FD}}(E_i(k)-\mu_i^*) +  f_{\mathrm{FD}} (E_i(k)+\mu_i^*) \right],
\end{eqnarray}
where $(2J_i+1)$ denotes the spin degeneracy factor,
$E_i=\sqrt{k^2+M^{*2}_i}$ is the single-particle energy. Interactions
enter via the effective masses $M^*_i$ and effective chemical
potentials $\mu^*_i$. The potential term is given by
\begin{eqnarray}
\label{eq:Z_mesons}
\frac{T}{V}\ln~{\cal Z}_m = 
  -\frac12 m_{\sigma}^2\bar\sigma^2
  + \frac12 m_{\omega}^2\bar\omega^2
  + \frac12 m_{\rho}^2\bar\rho^2 
  + \frac12 m_{\phi}^2\bar\phi^2~.
\label{eq:zm}
\end{eqnarray}
The values of the mean-fields having the quantum numbers of the
corresponding mesons, are determined by
\begin{subequations}
\begin{eqnarray}
  m_{\sigma}^2 \bar \sigma=\sum_{i} g_{\sigma i} n_i^s, \\
  m_{\omega}^2 \bar \omega=\sum_{i} g_{\omega i} n_i,\\
  m_{\phi}^2 \bar \phi=\sum_{i} g_{\phi i} n_i, \\
  m_{\rho}^2 \bar \rho=\sum_{i} g_{\rho i} t_{3i} n_i,
\end{eqnarray}
\end{subequations}
 {where} $t_{3i}$  {is} the third
component of isospin of baryon $i$, $n_i^s$ and $n_i$ are  {the} 
scalar and the number density. 
 {These} are given by
\begin{align}
  n_i^s&= \frac{1}{\pi^2} \int \frac{k^2 M^*_i}{E_i(k)}
  \left[f_{\mathrm{FD}}(E_i(k)-\mu^*_i)+ f_{\mathrm{FD}}(E_i(k)+\mu^*_i)  \right] dk, \\
  n_i&= \frac{1}{\pi^2} \int k^2\left[f_{\mathrm{FD}}(E_i(k)-\mu^*_i)- f_{\mathrm{FD}}(E_i(k)+\mu^*_i)  \right] dk.
\end{align}
Note that Eq.~(\ref{eq:zm}) does not contain mesonic
self-interactions, as is the case of many DFT models. Here, we employ
instead a model belonging to the subclass of DFTs which allow for
density-dependent couplings,
$g_{{\rm m,B}}(n_B)=g_{{\rm m,B}}(n_s) h_{\rm m} (x)$ with
$x=n_B/n_s$, where $n_s$ is the nuclear saturation density. We will
further assume that the couplings of mesons to hyperons and
$\Delta$-resonances have the same density dependence as those to
nucleons. The effective chemical potentials are given by
\begin{equation}
  \mu^*_i=\mu_i -g_{\omega i} \bar \omega -g_{\rho i} t_{3 i} \bar \rho -g_{\phi i} \bar \phi- \Sigma_0^R,
  \end{equation}
where
\begin{eqnarray}
  \Sigma_0^R=\sum_i \left(
  \frac{\partial g_{\omega i}}{\partial n_i} \bar \omega n_i+
  t_{3i} \frac{\partial g_{\rho i}}{\partial n_i} \bar \rho n_i+
    \frac{\partial g_{\phi i}}{\partial n_i} \bar \phi n_i-
     \frac{ \partial g_{\sigma i}}{\partial n_i} \bar \sigma n_i^s
  \right),
\end{eqnarray}
is the rearrangement term.
The effective baryon masses depend on the scalar mean field 
according to
\begin{equation}
  M_i^*=M_i-g_{\sigma i} \bar \sigma.
  \end{equation}

From the partition function one obtains pressure, entropy
density, energy density, and particle number densities in a standard fashion
\begin{eqnarray}
\label{eq:Thermo1}
  P&=&\frac{T}{V} \ln {\cal Z}, \\
\label{eq:Thermo2}
  s&=& \frac{1}{V}\frac{\partial \left( T \ln {\cal
       Z}\right)}{\partial T}\Big\vert_{V,\{\mu_i\}},\\
\label{eq:Thermo3}
  e&=&-\frac{T}{V} \ln {\cal Z}+\sum_i \mu_i n_i+T s,\\
\label{eq:Thermo4}
  n_i&=&\frac1V \frac{\partial \left( T \ln {\cal Z}\right)}{\partial \mu_i}\Big\vert_{V,T}.
\end{eqnarray}
Note that each component contributes additively to the net
thermodynamic quantity of interest, which is easy to see by
substituting Eqs.~\eqref{eq:Z_leptons}-\eqref{eq:Z_baryons} in
Eqs.~\eqref{eq:Thermo1}-\eqref{eq:Thermo3}.

In this work, we use the density-dependent DDME2 parameterisation for
the nucleonic sector \citep{Lalazissis_PRC_2005} and its extension to
the hyperonic sector by \cite{Fortin_PRC_2016}, which fixes the
hyperonic couplings in the vector meson sector to the values implied
by the $SU(6)$ symmetric quark model and adjusts the scalar couplings
to the depth of the hyperon potentials in nuclear matter at
saturation. Alternative extensions of the DDME2 model to the baryonic
octet have been carried out and applied to compact stars elsewhere
\citep{Colucci_PRC_2013,Dalen2014,Fortin2017,Li_PLB_2018}. The differences
between the models reside in the choice of the different hyperonic
couplings, which are not well constrained.  In addition to hyperons,
we include the {$\Delta(1232)$ resonance states of the baryon
  $J^{3/2}$-decouplet,} which were brought {recently} into focus in
the context of compact stars by several groups
\citep{Chen_PRC_2007,Drago_PRC_2014,Cai_PRC_2015,Zhu_PRC_2016,Sahoo_PRC_2018,Kolomeitsev_NPA_2017,Li_PLB_2018,Li2020PhRvD,Li2019ApJ,Ribes_2019}.
\begin{table}
  \begin{tabular}{lcccccc}
    \hline 
    Model & $n_{\rm s}$ & $E_{\rm s}$ & $K_{\rm inf}$  & $J$  & $L$  & $K_{\rm sym}$ \\
    &  [fm$^{-3}$] & [MeV] & [MeV] & [MeV]& [MeV] & [MeV] \\
    \hline
    DDME2 & 0.152 & -16.1 & 250.9 & 32.3 & 51.2 & -87.1 \\
 \hline
  \end{tabular}
  \caption{Key nuclear matter properties of the DDME2 
    model \citep{Lalazissis_PRC_2005}: the binding energy per nucleon ($E_{\rm s}$)
    and compression modulus ($K_{\rm inf}$)
    of symmetric nuclear matter at saturation density ($n_{\rm s}$) 
    together with the symmetry energy ($J$), its slope ($L$)
    and curvature ($K_{\rm sym}$).
  }
  \label{tab:snm}
\end{table}

Our choice of the nucleonic DDME2 EoS is motivated by the following
factors: i) the parameters of isospin symmetric ({\it i.e.}
equal numbers of protons and neutrons) nuclear matter
around saturation density are in good agreement with present
experimental constraints \citep{Lalazissis_PRC_2005} (see Table \ref{tab:snm});
ii) the properties of atomic nuclei, such as binding energies, rms radii
of charge distribution, neutron skin thickness,
  quadrupole and hexadecupole moments of heavy and
superheavy nuclei, excitation energies of the iso-scalar giant monopole- and
iso-vector giant dipole-resonances in spherical nuclei,
are in good agreement with experimental values \citep{Lalazissis_PRC_2005};
iii) the energy per baryon of
low-density neutron matter predicted by the DDME2 functional is in
good agreement with that from {\it ab initio} calculations
\citep{Gandolfi_PRC_2012,Hebeler_ApJ_2013} [see Fig. 12 in
\citep{Fortin_PRC_2016} and Fig. 1 in \citep{Li_PhysRevC.100.015809}].  Point
(iii) implies that the model predicts a relatively low value for the
slope of the symmetry energy $L$, which lies within the domains
$40 \lesssim L \lesssim 62$ MeV \citep{Lattimer_EPJA_2014} or
$30 \lesssim L \lesssim 86$ MeV \citep{Oertel_RMP_2017} deduced from
experiments [see Fig.~13 in \cite{Fortin_PRC_2016}].  The curvature of
the symmetry energy falls in the intervals $K_{sym}=-111.8 \pm 71.3$
MeV~\citep{Mondal_PRC_2017}, $K_{sym} = -85^{+82}_{-70}$
MeV~\citep{dEtivaux2019} and $K_{sym}=-102 \pm 71$ MeV
\citep{Zimmerman_2020},  {which were} obtained from the analyses of
different nuclear and compact star properties.  {This implies that
  the DDME2 parameterisation has a reasonable behavior in the
  isovector channel.}

The extension of the DDME2 to the hyperonic sector by
\cite{Fortin_PRC_2016} --labelled  {hereafter}
``DDME2Y''-- assumes SU(6) flavor
symmetry for the vector meson-hyperonic couplings and adjusts the
couplings of the scalar $\sigma$-meson to hyperons to reproduce the
empirical depths of the hyperon potentials in symmetric nuclear matter
at saturation. The potential for particle $j$ in $k$-particle matter
is thereby defined via the effective masses and chemical potentials as
\begin{equation}
U_j^{(k)}(n_k) = M^*_j - M_j + \mu_j - \mu_j^*~.
\end{equation}
The DDME2Y {parameterisation uses} the values
$U_{\Lambda}^{(N)} \approx -28$ MeV, $U_{\Xi}^{(N)} \approx -14$ MeV,
$U_{\Sigma}^{(N)} \approx 30$ MeV {in isospin symmetric nuclear
  matter} \citep{Gal_RMP_2016}.  Coupling constants of the hyperons
$Y$ to the meson fields are customarily expressed in terms of the
coupling constants of the nucleons $N$ to the meson fields,
$x_{m,Y}=g_{m,Y}/g_{m,N}$, where $m\in \sigma, \rho, \omega$,
etc. labels the meson.  Adopting this convention, DDME2Y
{parameterisation} is {defined} by the following coupling constants:
$x_{\sigma \Lambda}=0.615$, $x_{\sigma \Xi}$=0.3225,
$x_{\sigma \Sigma}=0.47$, $x_{\omega \Lambda}=2/3$,
$x_{\omega \Xi}=1/3$, $x_{\omega \Sigma}=2/3$, $x_{\rho \Lambda}=0$,
$x_{\rho \Xi}=1$, $x_{\rho \Sigma}=2$.  The coupling constants of the
hidden strangeness meson $\phi$ to the baryons are $g_{\phi N}$=0,
$g_{\phi \Lambda}=-\sqrt{2}/3 g_{\omega N}$,
$g_{\phi \Xi}=-2\sqrt{2}/3 g_{\omega N}$,
$g_{\phi \Sigma}=-\sqrt{2}/3 g_{\omega N}$.

The DDME2Y parameterisation above fulfills not only the existing
constraints from terrestrial experiments and {\it ab initio}
calculations, but it is in agreement with existing compact star
observations {as well} (see Section \ref{sec:NS} and Table
\ref{tab:NS}).  Specifically, the predicted value of the {compact
  star} maximum mass exceeds {the observational lower bound on the
  maximum mass of a compact star} $2M_{\odot}$
\citep{Demorest_2010,Antoniadis_2013,Arzoumanian_2018}.  Furthermore,
the predicted value of the radius of the canonical $1.4M_{\odot}$
compact star is in agreement with the recent inferences
$R= 13.02^{+1.24}_{-1.06}$~km~\citep{Miller_2019} and
$R= 12.71^{+1.14}_{-1.19}$ km~\citep{Riley_2019} from the data
obtained by the NICER mission.
The tidal deformability for a $1.4M_{\odot}$ compact star is
$\Lambda_{1.4} = 712$.
This value lies outside the range  $\Lambda_{1.4} = 190^{+390}_{-120}$ \citep{Abbott_2018}
and at the upper limit of the interval $300^{+420}_{-230}$ \citep{Abbott_PRX_2019}
extracted, at a 90\% confidence level, from the analysis of the
GW170817 event.
Note that the values of tidal deformability in \cite{Abbott_2018} have been
obtained assuming that both compact objects are
NSs obeying a common equation of state; on the other hand, the 
values in \cite{Abbott_PRX_2019} have been obtained,
as in the initial analysis of GW170817 event \citep{Abbott_2017}, by making
minimal assumptions about the nature of the compact
objects and allowing the tidal deformability of
each object to vary independently.

The masses of $\Delta$-resonances, which form an isospin quadruplet,
lie between those of $\Sigma$ and $\Xi$ hyperons, {therefore} they are
expected to nucleate in dense {stellar} matter {according to} the same
energetic arguments employed {for the} nucleation of hyperons
\citep{Glendenning_ApJ_1985}.  {While in the vacuum} $\Delta$s are
broad resonances which decay into nucleons with emission of a pion, in
stellar matter they are thought to be stabilised by the
{Pauli-blocking of the final} nucleon states.  Apart from the
narrowing the quasi-particle width of the $\Delta$s, matter effects may
shift the quasi-particle energy to larger values
\citep{Sawyer1972,Ouellette2001}, which would suppress the $\Delta$
degrees of freedom.    {Below, we will assume that the $\Delta$s
retain their vacuum masses and have negligible width, as has been done
in the recent
literature~\citep{Chen_PRC_2007,Drago_PRC_2014,Cai_PRC_2015,Zhu_PRC_2016,Sahoo_PRC_2018,Kolomeitsev_NPA_2017,Li_PLB_2018,Li2020PhRvD,Li2019ApJ,Ribes_2019}}.

 {The information about nucleon-$\Delta$ interaction} is extracted
from pion-nucleus scattering and pion photo-production
\citep{Nakamura_PRC_2010}, electron scattering on nuclei
\citep{Koch_NPA_1985} and electromagnetic excitations of the
$\Delta$-baryons \citep{Wehrberger_NPA_1989}.  
As reviewed by \cite{Drago_PRC_2014} and \cite{Kolomeitsev_NPA_2017}
(a) the potential of the $\Delta$ in the nuclear medium is slightly
more attractive than the nucleon potential
$-30~{\rm MeV}+U^{(N)}_N \lesssim U^{(N)}_{\Delta} \lesssim
U^{(N)}_N$,
which translates in values of $x_{\sigma \Delta}$ slightly larger than
1, (b) $0 \lesssim x_{\sigma \Delta}- x_{\omega \Delta} \lesssim 0.2$
and (c) no experimental constraints exist for the value of
$x_{\rho \Delta}$.

Since there remain large uncertainties on the values of the $\Delta$
couplings, they are commonly varied in a certain plausible
range.  Previous works employed the ranges
$0.85 \lesssim x_{\sigma \Delta} \lesssim 1.15$,
$0.6 \lesssim x_{\omega \Delta} \lesssim 1.2$, and
$0.5 \lesssim x_{\rho \Delta} \lesssim
3$~\citep{Chen_PRC_2007,Drago_PRC_2014,Cai_PRC_2015,Zhu_PRC_2016,Sahoo_PRC_2018,Kolomeitsev_NPA_2017,Li_PLB_2018,Li2020PhRvD,Li2019ApJ,Ribes_2019}.
 {Furthermore,} it was shown that:
\begin{itemize}

\item small values of $x_{\sigma \Delta}$ (which lead to small values $U_{\Delta}^{(N)}$)
  result in larger values of compact star radii
  \citep{Zhu_PRC_2016,Kolomeitsev_NPA_2017,Spinella-PhD,Li_PLB_2018};
  the effect on the maximum mass is small and dependent on the high density
  part of the nucleonic EoS, 
  
\item small values of $x_{\omega \Delta}$ and $x_{\rho \Delta}$ imply
  lower values of  {compact star} radii and maximum masses
  \citep{Drago_PRC_2014,Cai_PRC_2015,Sahoo_PRC_2018,Li2020PhRvD,Li2019ApJ,Ribes_2019},

\item the threshold density for the onset of $\Delta$ is correlated with
  $x_{\sigma \Delta}$ \citep{Zhu_PRC_2016,Kolomeitsev_NPA_2017,Spinella-PhD,Li_PLB_2018},
  $x_{\rho \Delta}$
  \citep{Cai_PRC_2015,Zhu_PRC_2016,Sahoo_PRC_2018},
  $x_{\omega \Delta}$ \citep{Ribes_2019} and the $\Delta$-effective mass
  \citep{Cai_PRC_2015,Sahoo_PRC_2018},

\item The effective mass of $\Delta$s significantly impacts  {compact star} radii and
  maximum masses \citep{Cai_PRC_2015,Sahoo_PRC_2018},

\item the threshold density for the onset of $\Delta$ strongly depends
  on the slope of the symmetry energy at saturation
  \citep{Drago_PRC_2014,Cai_PRC_2015}.  \cite{Cai_PRC_2015}  {also
    showed that it is less sensitive} to the other parameters of
  symmetric saturated nuclear matter.
\end{itemize}

The appearance of $\Delta$ resonances in hot stellar
matter with fixed lepton fraction was investigated only recently by
\cite{Malfatti_PRC_2019}. It was found that: (i) the lower the lepton
fraction, the higher the $\Delta$ abundances; (ii) at high enough
temperatures and densities the four isobars are populated in addition
to all hyperonic degrees of freedom; (iii) the most abundant of the
$\Delta$-isobars is $\Delta^-$.

In this work, we use the following values of the couplings of
mesons to $\Delta$s: $x_{\sigma \Delta}=1.1$, which corresponds to a
$\Delta$ potential at the saturation $U_{\Delta}^{(N)} \approx -83$
MeV, $x_{\omega \Delta}=1.1$, and $x_{\rho \Delta}=1.0$   {and} $x_{\phi \Delta}=0$. 
In the following this model will be referred to as DDME2Y$\Delta$.

We assume strangeness changing weak equilibrium 
leading to the following equilibrium conditions,
\begin{eqnarray}
\label{eq:equilib1}
  \mu_{\Lambda}=\mu_{\Sigma^0}=\mu_{\Xi^0}=\mu_{\Delta^0}=\mu_n=\mu_B;\\
\label{eq:equilib2}
  \mu_{\Sigma^-}=\mu_{\Xi^-}=\mu_{\Delta^-}=\mu_B-\mu_Q;\\
\label{eq:equilib3}
  \mu_{\Sigma^+}=\mu_{\Delta^+}=\mu_B+\mu_Q;\\
\label{eq:equilib4}
  \mu_{\Delta^{++}}=\mu_B+2\mu_Q,
\end{eqnarray}
 {where $\mu_B$ is the baryon number chemical potential and
  $\mu_Q=\mu_p-\mu_n$ is the charge chemical potential.} The
equilibrium conditions, Eqs.~\eqref{eq:equilib1}-\eqref{eq:equilib4},
together with the total baryonic charge
$n_p+n_{\Sigma^+}+2n_{\Delta^{++}}+n_{\Delta^{+}}-(n_{\Sigma^-}+n_{\Xi^-}+n_{\Delta^-})=n_Q$
determine the composition of baryonic matter for a given
$(n_B, T, Y_Q = n_Q/n_B)$.  {The requirement of} global electrical
charge neutrality of stellar matter then fixes the charged lepton
density $Y_Q = Y_e + Y_\mu$ where $Y_e = (n_{e^-} - n_{e^+})/n_B$ and
$Y_\mu = (n_{\mu^-} - n_{\mu^+})/n_B$. For free-streaming neutrinos
$\mu_e=\mu_\mu=-\mu_Q$, whereas for trapped neutrinos
$\mu_{e/\mu}=\mu_{L,e/\mu}-\mu_Q$, where $\mu_{L,e/\mu}$ denotes the
(electron/muon) lepton number chemical potential.  The lepton $Y_{L,e/\mu}$
  fractions, which are conserved separately,  are then defined via
the total lepton number density divided by $n_B$.  
Throughout this paper ``$\beta$-equilibrium"
refers to $\beta$-equilibrated matter which, in addition, 
{\it is transparent to neutrinos}. This means that the corresponding
lepton number chemical potentials vanish
$\mu_{L,e} = \mu_{L,\mu} = 0$.

At densities below nuclear saturation density and not too high
temperature, the matter becomes unstable towards density fluctuations,
because of the competition between nuclear and Coulomb
interactions. As a consequence, a large variety of clusters are
formed, all of which are in chemical and thermal equilibrium with the
unbound baryons. If the temperature is high enough, hyperons and
$\Delta$s are in principle expected to nucleate within the clusters
and unbound components as well, but for simplicity, we neglect this
possibility here. The theoretical framework suitable under these
conditions is the Nuclear Statistical Equilibrium (NSE)
\citep{Hempel_NPA_2010,Raduta_PRC_2010,Raduta_PRC_2015}.  Interactions
among unbound particles and clusters are usually accounted for in the
excluded volume approximation, while those in the homogeneous matter
within a chosen mean-field approach.  The transition between the
inhomogeneous and homogeneous phases is in principle realized by
minimizing, for equal values of the intensive thermodynamic
observables, the relevant thermodynamic potential. The EoS used in our
work are obtained by smoothly merging the uniform matter EoS to the
NSE model HS(DD2) \citep{Hempel_NPA_2010} for inhomogeneous
matter. The latter is publicly available on the Compose
database\footnote{\url{https://compose.obspm.fr/}}~\citep{Typel_compose}.
In principle, the transition density depends on the EoS, temperature and
  charge fraction, see e.g. the discussion in
  \citep{Ducoin:2008xs,Ducoin:2006td,Pais:2010dp}, but for simplicity,
the matching is performed here at a fixed transition density
$n_t=n_s/2$. Inhomogeneous and homogeneous matter are considered
  at the same $S/A$ and $Y_{L,e}/\mu_{L,e}$. The similarity between
the effective interactions of DDME2 \citep{Lalazissis_PRC_2005} and
DD2 \citep{Typel_PRC_2010} leads to a coherent treatment of the EoS
over the whole density regime and the fixed transition density only
induces very small thermodynamic inconsistencies with little
  impact on the EoS and the global star properties studied here.

\begin{table*}
  \begin{tabular}{lcclcclcclcccc}
    \hline 
    Model &     $M_{G,{\rm max}}$ & $n_{c,{\rm max}}$  &
    $Y_1$ & $n_{Y_1}$ & $M_{Y_1}$ & $Y_2$ & $n_{Y_2}$  & $M_{Y_2}$ & $Y_3$ & $n_{Y_3}$  & $M_{Y_3}$ &
    $R_{1.44M_{\odot}}$ & $\Lambda_{1.4}$ \\
    &     $(M_{\odot})$ &  $({\rm fm}^{-3})$ &
    & $({\rm fm}^{-3})$ & $(M_{\odot})$ & & $({\rm fm}^{-3})$ & $(M_{\odot})$ & & $({\rm fm}^{-3})$ &
    $(M_{\odot})$ & (km) & \\
    \hline
    DDME2Y         & 2.113 & 0.93 &   
    $\Lambda$ & 0.34 &  1.39 & $\Xi^-$ & 0.37 & 1.54 & $\Sigma^-$ & 0.39 & 1.60 & 13.25 & 712 \\
    DDME2Y$\Delta$ & 2.111 & 0.96 &
    $\Delta^-$ & 0.28 &  0.96 &  $\Lambda$ & 0.36 & 1.33 &  $\Xi^-$ & 0.52 & 1.82  & 13.09 & 653 \\
    \hline
  \end{tabular}
  \caption{    Properties of non-rotating spherically symmetric cold
    $\beta$-equilibrated,   {neutrino-transparent,  compact stars}  {based on }
       the EoS models considered in this work.
    $n_{c,{\rm max}}$ stands for the central baryon number 
    density of the maximum gravitational mass
    ($M_{G,{\rm max}}$) configuration.
    Columns 4, 7 and 10 specify the heavy baryon species that nucleate in stable stars.
    $n_i$ represents the threshold density at which the species $i$ is produced, while
    $M_i$ gives the corresponding gravitational mass of the star.
    $R_{1.44M_{\odot}}$ indicates the radius of a fiducial $1.44M_{\odot}$ star;
    $\Lambda_{1.4}$ represents the tidal deformability of a canonical $1.4M_{\odot}$ star.
  }
  \label{tab:NS}
  \end{table*}

\section{Equation of state and composition}
\label{sec:NS}

A PNS is born in the aftermath of a successful supernova explosion,
when the stellar remnant and the expanding ejecta get gravitationally
decoupled.  The evolutionary epoch during which the remnant changes
from a hot and lepton-rich PNS to a cold and deleptonized compact star
lasts for several tens of seconds and consists of two major
evolutionary stages \citep{Prakash_1997,Pons_ApJ_1999}: the
deleptonization stage and the cooling stage.  The deleptonization
stage is characterized by a gradual decrease of the net lepton and
proton fractions and the heat-up of the core, due to the diffusion of
trapped electron neutrinos from the central region outward.  The
cooling stage is characterized by a simultaneous decrease of both
entropy and lepton content.   {The} structure and composition of the PNS
during this epoch will be investigated here in a schematic way,
assuming entropy per baryon and lepton fraction with typical values
\citep{Prakash_1997,Pons_ApJ_1999}: ($S/A=1$, $Y_{L,e}=0.4$), ($S/A=2$,
$Y_{L,e}=0.2$), ($S/A=1$, $\mu_{L,e/\mu}=0$) and ($S/A=0$, $\mu_{L,e/\mu}=0$).

\subsection{Composition}
\label{ssec:compo}

\begin{figure}
  \begin{center}
    \includegraphics[angle=0, width=0.99\columnwidth]{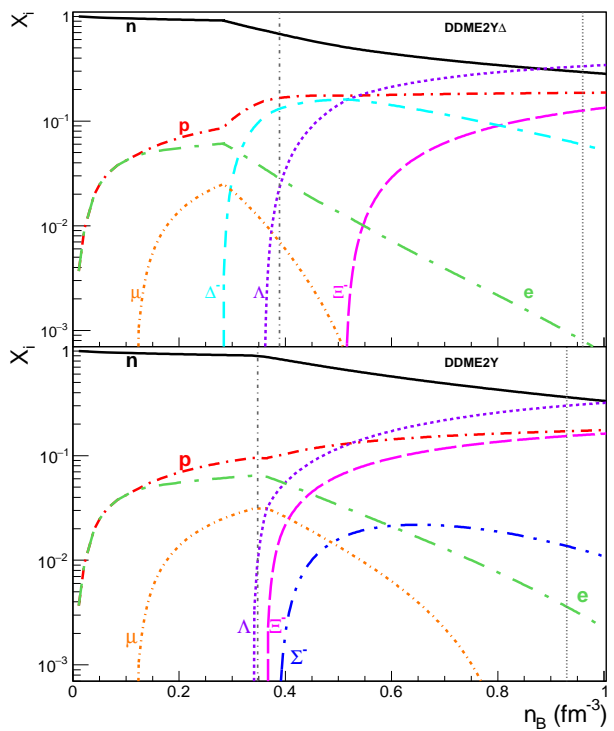}
 \end{center}
 \caption{Relative abundances in $\beta$-equilibrated,
   neutrino-transparent, cold compact star matter as predicted by
   DDME2Y (bottom panel) and DDME2Y$\Delta$ (top panel) models as
     function of baryon number density.  Note that the nucleation of
   $\Delta^-$ resonance leads to a suppression of $\Sigma^-$
   abundance.  Thin vertical lines mark the central baryon
     number densities corresponding to a $1.44M_{\odot}$ star
     (dot-dashed) and the maximum mass star (dotted), respectively.
 }
  \label{fig:Xi_T=0}
\end{figure}

\begin{figure}
  \begin{center}
    \includegraphics[angle=0, width=0.99\columnwidth]{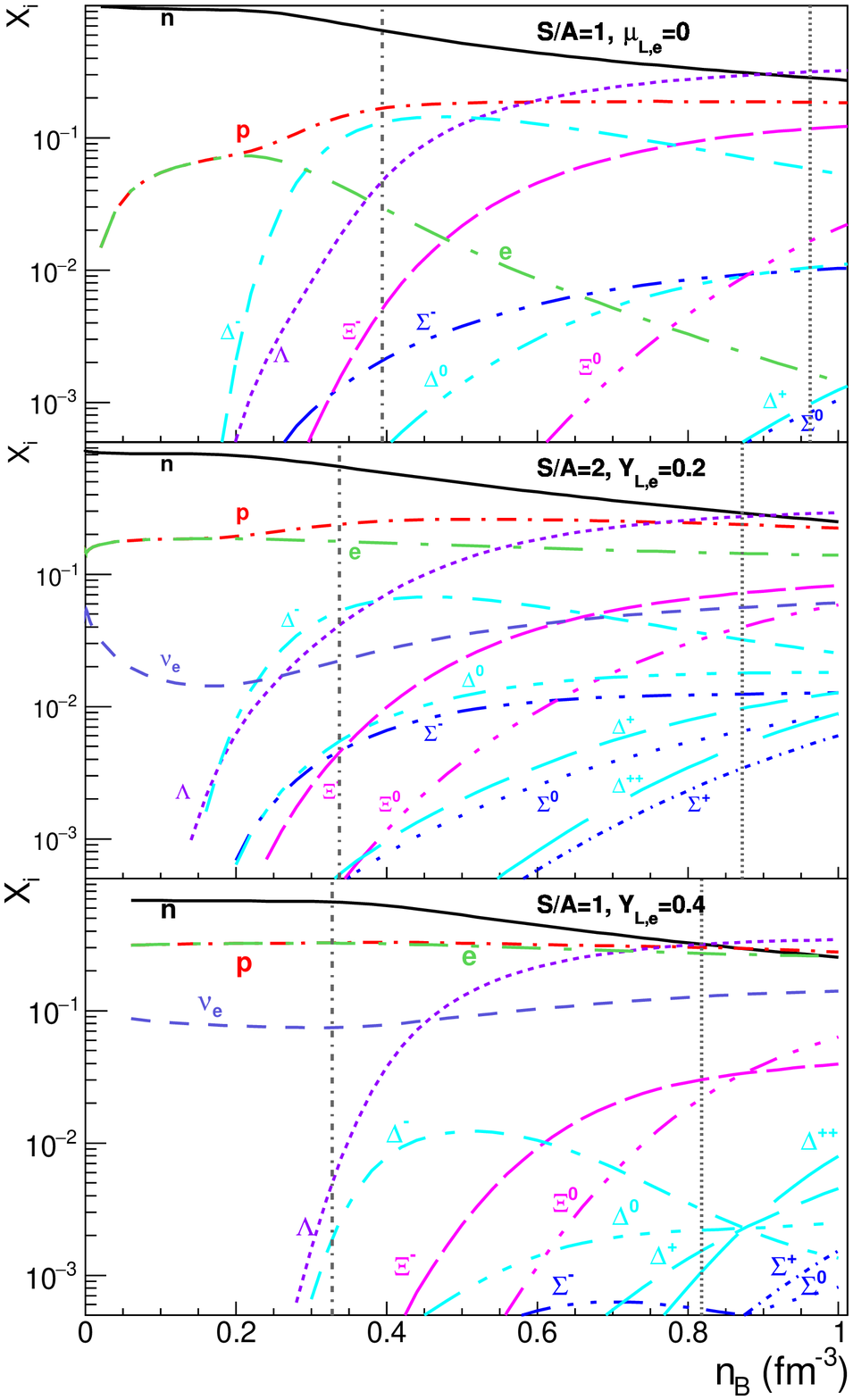}
  \end{center}
  \caption{
    Relative particle abundances in hot $NY \Delta$ matter
    with ($S/A=1$, $Y_{L,e}=0.4$) (bottom), ($S/A=2$, $Y_{L,e}=0.2$) (middle) and 
    ($S/A=1$, $\mu_{L,e}=0$) (top),
    as predicted by the DDME2Y$\Delta$ model. Muons are not considered. 
   Thin vertical lines mark the central baryon number densities corresponding
   to a $1.44M_{\odot}$ star (dot-dashed) and a  maximum mass star (dotted), respectively.
  }
  \label{fig:Xi_T}
\end{figure}


Fig. \ref{fig:Xi_T=0} shows relative particle abundances as a function
of baryon number density in cold $\beta$-equilibrated {compact star}
matter.  Results corresponding to baryonic matter composed of
$NY\Delta$ (top panel) are compared with those corresponding to $NY$
(bottom panel).  In both cases, the net charge neutrality is
guaranteed by electrons and muons.  In the case of $NY$ matter, the
only non-nucleonic baryonic degrees of freedom that nucleate in stable
stars are $\Lambda$, $\Xi^-$ and $\Sigma^-$, as already discussed
elsewhere \citep{Fortin_PRC_2016,Raduta_MNRAS_2018}.  This can be
explained by the large negative charge chemical potential in matter
featuring low charge fractions favoring negatively charged
particles. Under the considered conditions, the $\Lambda$-hyperons
remain nevertheless the most abundant non-nucleonic species. Their
lower mass compared with those of $\Sigma^-$- and $\Xi^-$-hyperons and
more attractive potential in nuclear matter
compensate the effect of the charge chemical potential.

In the case of $NY\Delta$ matter, $\Delta^-$ is the first heavy baryon
to appear.  Its onset strongly affects the hyperonic abundances: the
threshold densities for the appearance of $\Lambda$ and $\Xi^-$
hyperons are shifted to higher densities and the $\Sigma^-$ hyperon is
completely suppressed.  Nucleation of $\Delta^-$s also modifies the
neutron and proton abundances: neutron (proton) abundance in
$NY\Delta$ matter is smaller (larger) than in $NY$ matter.  The
proton fraction remains below the threshold for the nucleonic
dUrca process. \footnote{Note that in several
  models, {\it e.g.} SWL \citep{Spinella-PhD} and FSU2H
  \citep{Tolos2017a,Tolos2017b},  {the} nucleation of $\Delta^-$
  opens up nucleonic dUrca  process or shifts its threshold to
  much lower densities.}  Recall
that, due to the relatively low $L$-value, neither purely nucleonic
nor hyperon admixed compact stars based on
DDME2(Y) EoS allow for a nucleonic dUrca process 
\citep{Fortin_PRC_2016}.  Finally,  we observe
  that the onset of $\Delta^-$ leads to a fast drop of lepton
  densities, as they compensate for the charge of leptons and are more
  energetically favorable than the electrons and muons. 

The properties of non-rotating spherically-symmetric cold
$\beta$-equilibrated compact stars based on
DDME2Y($\Delta$) models are summarized in Table
\ref{tab:NS}.  {Listed are} the maximum gravitational mass
$M_{G,{\rm max}}$, the corresponding central baryon number density
$n_{c,{\rm max}}$, the threshold densities for nucleation of heavy
baryons and the corresponding minimal  {compact star} masses, the
radius of a  {compact star} with a canonical mass of
$1.44 M_\odot$.  We note that (i) with $M_{G,{\rm max}}=2.1M_{\odot}$,
DDME2Y and DDME2Y$\Delta$ models both fulfil the $2M_{\odot}$
constraint on the lower limit of maximum mass of a compact star
\citep{Demorest_2010,Antoniadis_2013,Arzoumanian_2018}; (ii) the
radii, $R_{1.44M_{\odot}}^{(NY)}= 13.25$ km and
$R_{1.44M_{\odot}}^{(NY\Delta)}=13.09$ km, are in agreement with
recent NICER results giving $13.02^{+1.24}_{-1.06}$ km
\citep{Miller_2019} and, respectively, $12.71^{+1.14}_{-1.19}$ km
\citep{Riley_2019} for PSR J0030+0451 with a mass of
$M_G=1.44^{+0.15}_{-0.14}$ \citep{Miller_2019} and
$M_G=1.34^{+0.15}_{-0.16} M_{\odot}$ \citep{Riley_2019};
(iii) $\Lambda_{1.4}=653$ is in better agreement with \cite{Abbott_2018};
(iv) the
$1.44M_{\odot}$ stars will have a tiny fraction of $\Lambda$-hyperons,
and, within the DDME2Y$\Delta$ model, a  {core which is rich with
}$\Delta^-$ resonances.  

The composition of matter at non-zero temperature is modified because
of the thermal excitation of new degrees of freedom.  Possible
neutrino trapping additionally modifies the composition.
Fig. \ref{fig:Xi_T} illustrates, as a function of baryon number
density, the relative particle abundances of hot $NY \Delta$ matter
under thermodynamic conditions relevant for different stages in the
evolution of PNS \citep{Pons_ApJ_1999}.  Each panel corresponds to a
set of constant values of total entropy per baryon and electron lepton
fraction.  In all cases, a vanishing muon lepton fraction
$Y_{L,\mu}=0$ is assumed.  The bottom panel corresponds to a moment
shortly after core bounce, when the star is hot and lepton rich
($S/A=1$, $Y_{L,e}=0.4$); the middle panel corresponds to a later
time, when the star is partially deleptonized and hotter ($S/A=2$,
$Y_{L,e}=0.2$); finally the top panel corresponds to fully
deleptonized matter which cools down ($S/A=1$, $\mu_{L,e}=0$). 

At high enough temperatures and low-$Y_{L,e}$ values heavy baryons can
nucleate already at sub-saturation densities.  Moreover, the heavy
baryon fractions increase with temperature, and eventually, all allowed
particle degrees of freedom can be populated.  As visible from
Fig. \ref{fig:Xi_T}, the dependence on $n_B$ is not always monotonic,
due to the competition among different species.  Still, due to the
large negative charge chemical potential in matter with low charge fractions,
negatively charged baryons nucleate at lower densities and are more
abundant than their neutral and positive counterparts. The charge
chemical potential decreases with the charge fraction, such that the
effect becomes more pronounced at low values of $Y_{L,e}$. Thus,
$\Lambda$-hyperons are the most abundant non-nucleonic species, except
for low values of $Y_{L,e}$, where the $\Delta^-$ abundance can exceed
the $\Lambda$ one. Our results qualitatively agree with the findings
by \cite{Oertel_PRC_2012,Oertel_EPJA_2016}, where a large number of
EoS models with hyperons was considered, and by
\cite{Malfatti_PRC_2019} concerning the $\Delta$s.

\subsection{Equation of state}
\label{ssec:eos}

\begin{figure}
\begin{center}
\includegraphics[angle=0, width=0.99\columnwidth]{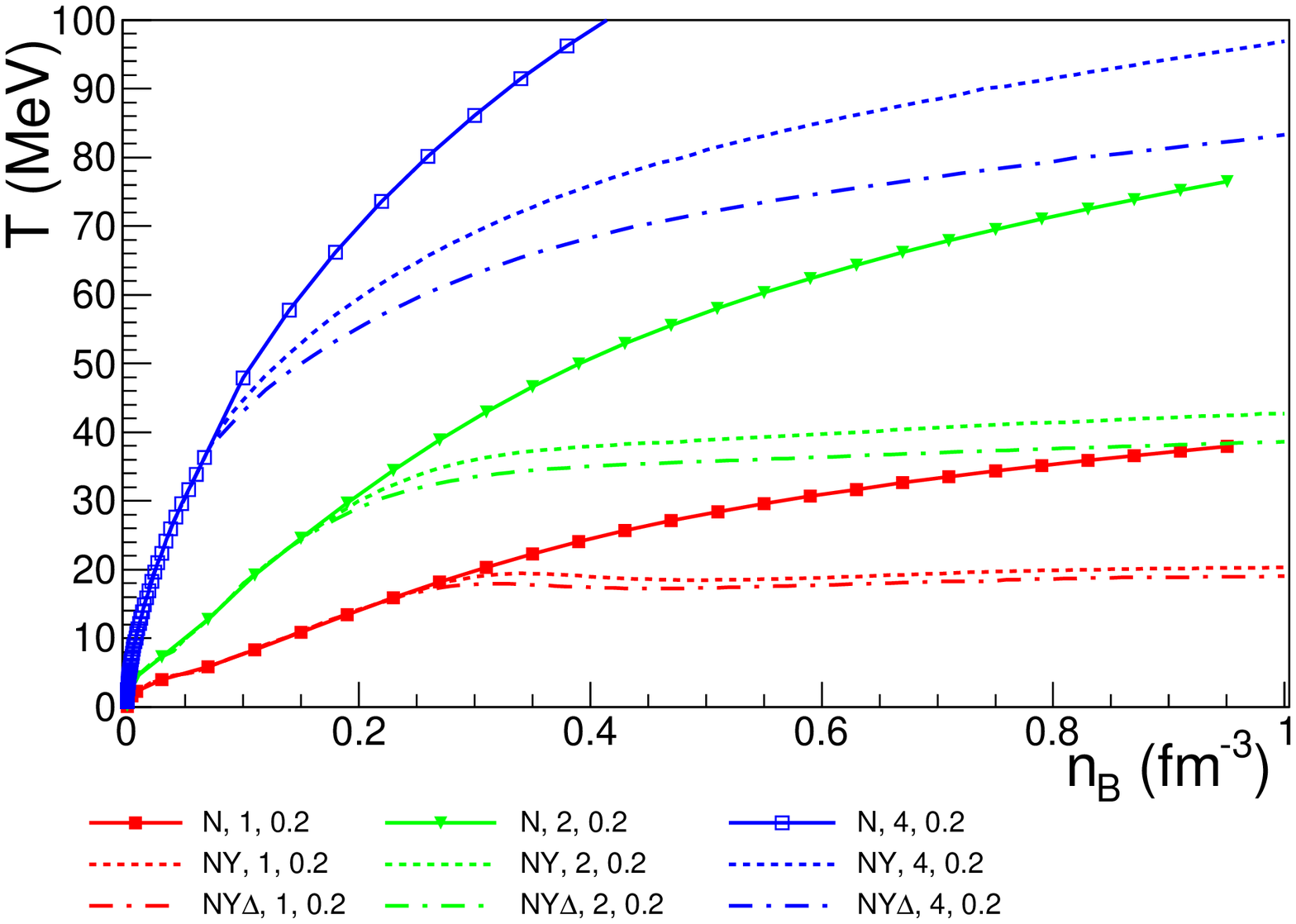}
\includegraphics[angle=0, width=0.99\columnwidth]{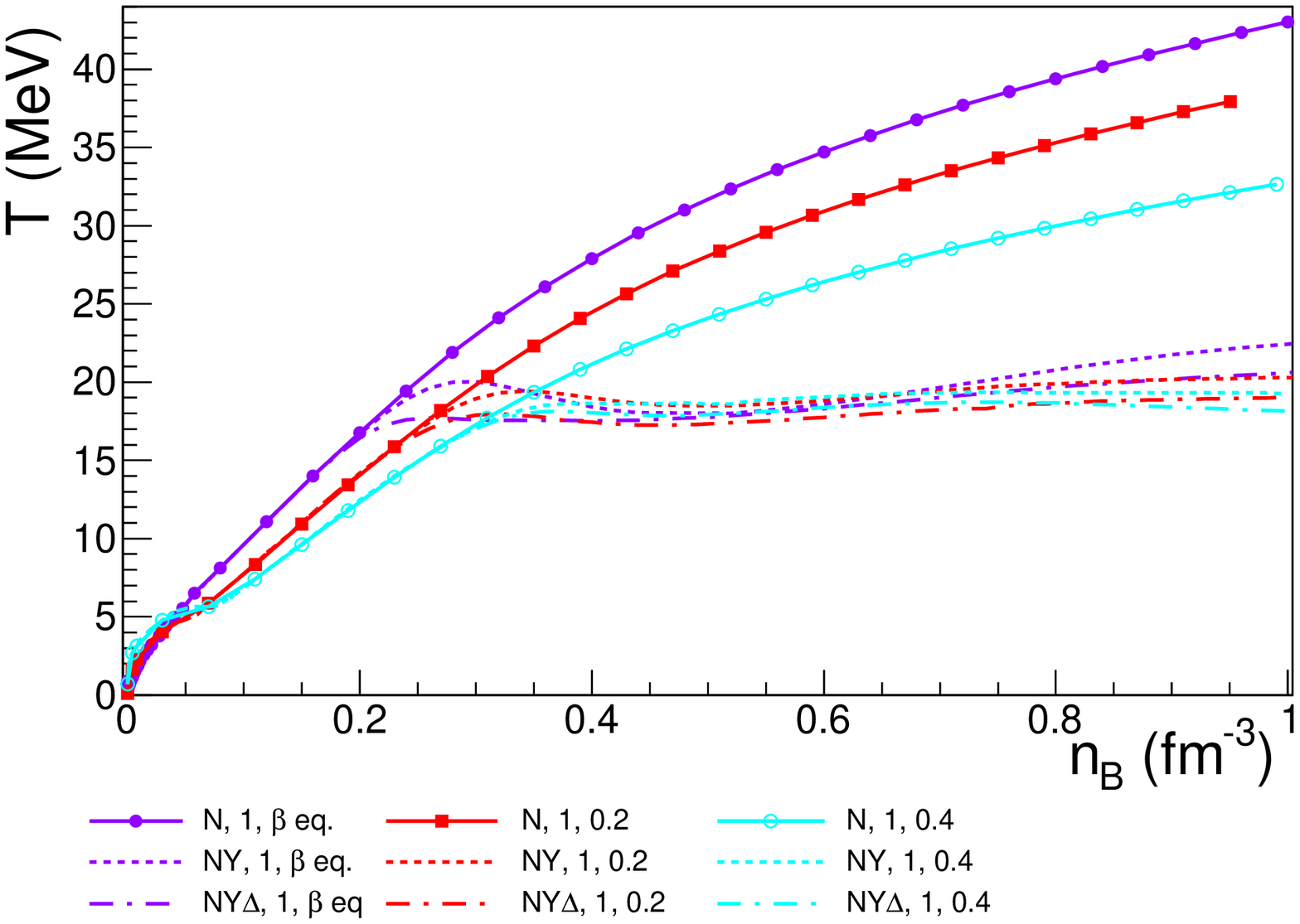}
\includegraphics[angle=0, width=0.99\columnwidth]{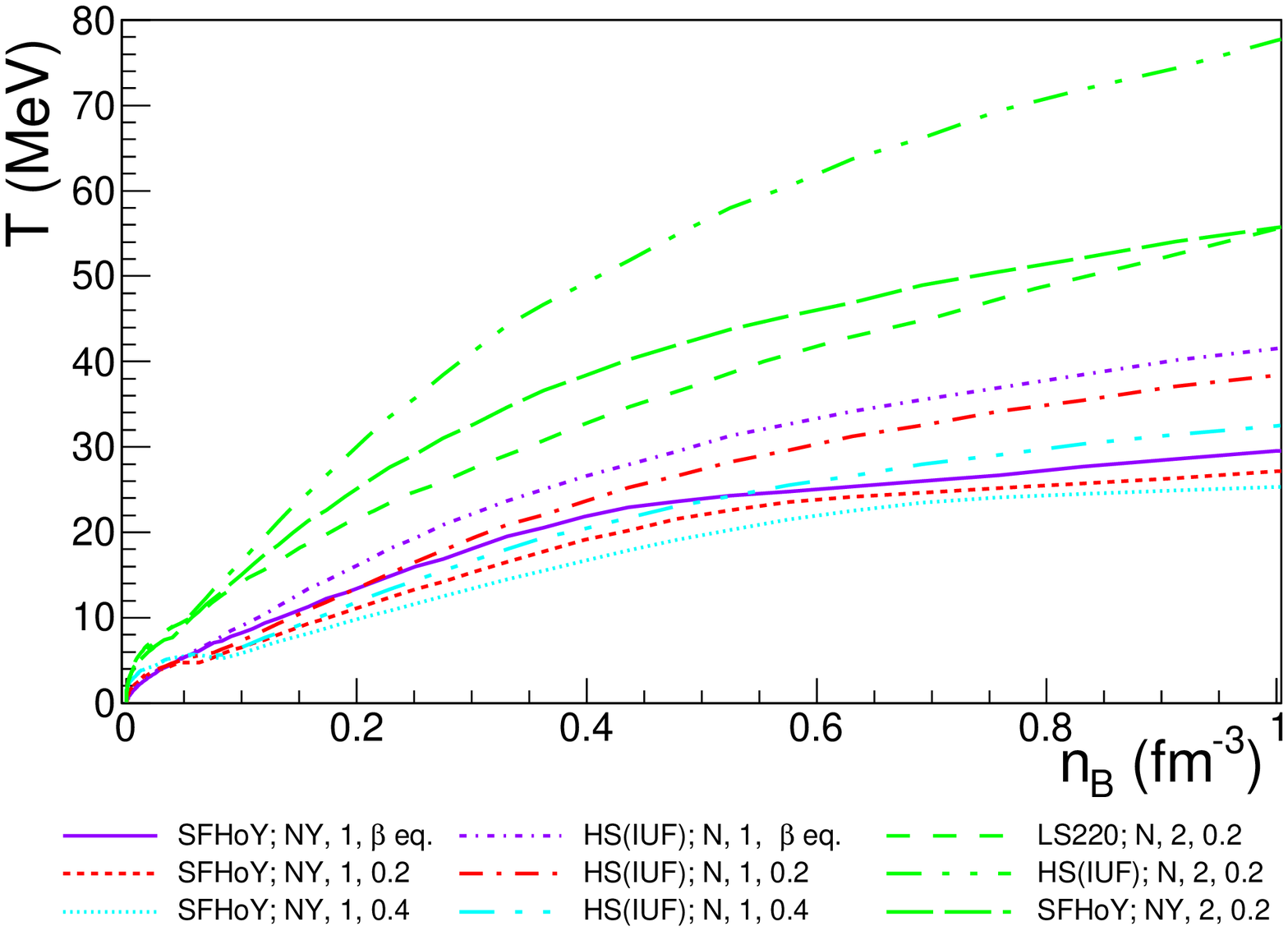}
\end{center}
\caption{Temperature versus baryon number density for hot star matter
  whose baryonic sector allows for nucleons ($N$), $NY$ and $NY\Delta$.
  Top panel: predictions of DDME2(Y$\Delta$) model for different values of
  $S/A$ and $Y_{L, e} = 0.2$, as given in the legend in the format $S/A,Y_{L,e}$.
  Middle panel: the same for $S/A = 1$ and varying $Y_{L,e}$ as well as
  $\mu_{L,e}=0$.
  Bottom panel: predictions of purely nucleonic EoS models
  LS220 \citep{LS_NPA_1991} and HS(IUF) \citep{Fischer2014} as well as
  hyperonic SFHoY EoS \citep{Fortin_PASA_2018}, for different values of
  $S/A,Y_{L,e}$.
  In all cases $Y_{L,\mu}=0$.
}
\label{fig:T-nB}
\end{figure}


Fig. \ref{fig:T-nB} shows the temperature as function of baryon number
density for matter composed of nucleons (N), nucleons and hyperons
($NY$), and nucleons, hyperons and $\Delta$ ($NY\Delta$), as predicted
by the DDME2, DDME2Y, and DDME2Y$\Delta$ model, respectively.  The top
panel compares different values of $S/A$ for the same $Y_{L,e} = 0.2$
whereas the middle panel compares different values of $Y_{L,e}$ for
the same $S/A = 1$.  For purely nucleonic matter, a strong increase of
temperature with density is observed over the entire density range,
whereas the increase is much less steep as soon as additional
particles appear.  Temperature can even decrease with the density over
a certain range, see {\it e.g.} the curves for $S/A = 1$
 {for} the DDME2Y and DDME2Y$\Delta$ models. The
change of slope is due to the sequential onset of heavy baryons (see
Fig. \ref{fig:Xi_T}).  A comparison between the results obtained for
$N$, $NY$ and, respectively, $NY\Delta$ matter proves that for fixed
values of $(n_B, Y_{L,e}, S/A)$, lower values of temperature are
reached in systems with more particle degrees of freedom.  This can be
explained by the fact that, at a given temperature, the entropy of a
system increases with the number of constituent particles.  The effect
was already discussed by \cite{Oertel_EPJA_2016}, who confronted the
behavior of matter composed of nucleons and hyperons to that of purely
nucleonic matter and, respectively, nucleons and $\Lambda$-hyperons.
 {They} also showed that, for fixed $n_B$ and $S/A$,
the uncertainties in the modeling of the nucleonic sector, especially
in the isovector channel, induce larger variations in temperature than
those arising from modeling hyperonic matter, see the results
corresponding to the purely nucleonic EoS of \cite{LS_NPA_1991},
labelled ``LS220'', the IUFSU~\citep{Fattoyev_PRC_2010} version of
\cite{Fischer2014} labeled ``HS(IUF)'' as well as the hyperonic EoS
``SFHoY'' \citep{Fortin_PASA_2018} at $S/A=2$ and
$Y_{L,e}=0.2$\footnote{These EoS are available on the Compose
  database.}, plotted in the bottom panel.

\begin{figure}
  \begin{center}
    \includegraphics[angle=0, width=0.99\columnwidth]{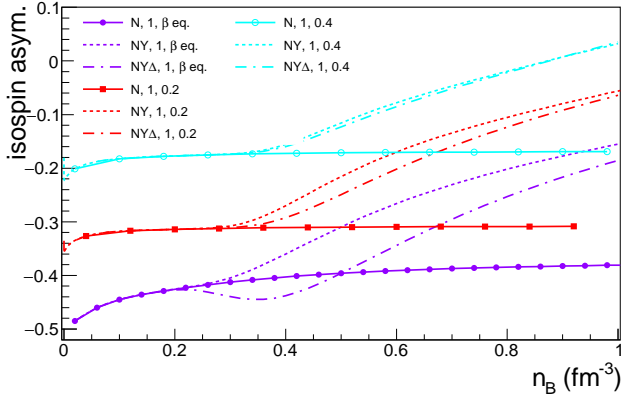}
    \end{center}
    \caption{ Isospin asymmetry $\sum_i t_{3,i}n_i$ as a function of
      particle number density for $\beta$-equilibrated,
      neutrino-transparent, matter and matter with constant (electron)
      lepton fractions at $S/A=1$, as predicted by DDME2, DDME2Y and
      DDME2Y$\Delta$ models.  }
  \label{fig:isoasym}
\end{figure}

The middle panel of Fig. \ref{fig:T-nB} demonstrates that, for given baryon
content of matter ({\it i.e.} $N$ vs. $NY$ vs. $NY\Delta$) and fixed
values of $S/A$ and $n_B$, the temperature depends on $Y_{L,e}$.  In
all cases, for a given $S/A$, the temperature decreases with
increasing $Y_{L,e}$.  The amplitude of this effect depends on the
available particle degrees of freedom via the overall isospin
asymmetry of matter,  {shown} in
Fig. \ref{fig:isoasym}.  The smaller this asymmetry, the smaller the
temperature variation as a function of $Y_{L,e}$. In particular, in
the presence of hyperons, isospin asymmetry becomes reduced for a
given $Y_{L,e}$, reflecting the fact that some neutrons are replaced
by $\Lambda$-hyperons, such that the temperature variation is weaker.

\begin{figure}
  \begin{center}
    \includegraphics[angle=0, width=0.99\columnwidth]{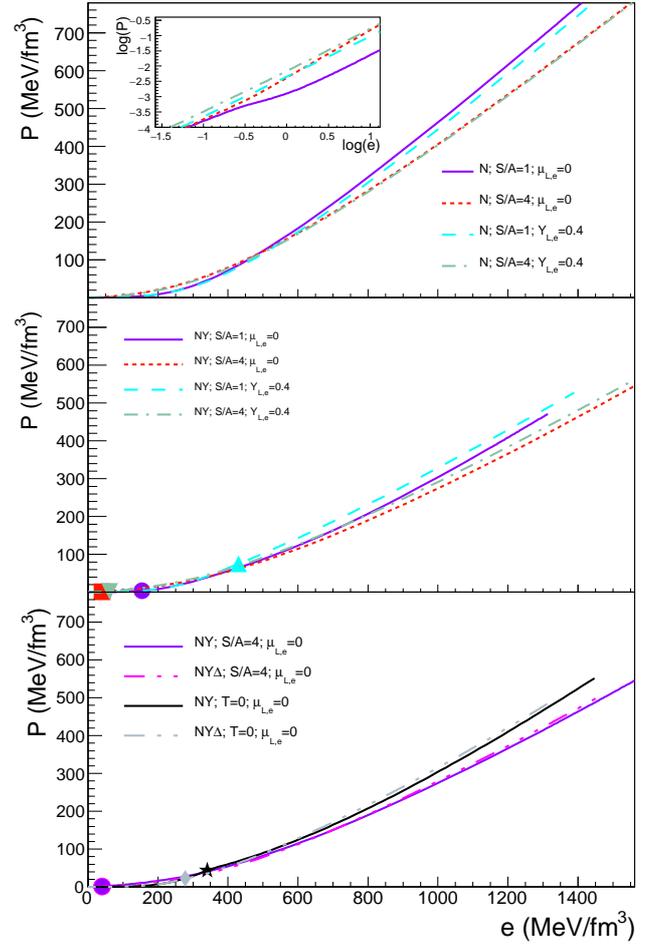}
    \end{center}
    \caption{EoS for different values of $S/A$ and $Y_{L,e}$ (or $\mu_{L,e}=0$), as given
      in the legend according to the DDME2($Y\Delta$).
      The inset in the top panel illustrates the behavior of the nucleonic matter EoS
        in the intermediate energy density range.
        The modifications of the EoS due to the
        onset of $\Delta$s are illustrated in the bottom panel for the case of
        $\beta$-equilibrated, neutrino-transparent matter.
        The onset of heavy baryons is marked, in each case, by a symbol.
      }
  \label{fig:Pe}
\end{figure}

Fig. \ref{fig:Pe} shows the dependence of total pressure on the total
energy density. Predictions for purely nucleonic matter (top) are
  confronted with those corresponding to matter composed of nucleons
  and hyperons (middle panel) for $S/A=1$ and 4, at $\mu_{L,e}=0$ and
  $Y_{L,e}=0.4$. The lower panel shows the impact of the onset of
  $\Delta$s at $\mu_{L,e} = 0$ and $S/A = 0$ and 4.
We find that i) for low energy densities $e \lesssim 300$
MeV/fm$^3$, the pressure increases with $(S/A)$, whereas the opposite
is is true at higher energy densities; ii) the dependence of
the pressure on $Y_{L,e}$ is complex; for purely nucleonic matter
  at high $e$, pressure is higher for lower $Y_{L,e}$-values, whereas
  the opposite is observed for matter containing hyperons; for
intermediate energy-densities lower $Y_{L,e}$-values lead to lower
values of pressure, no difference between nucleonic and hyperonic EoS
are observed here since hyperons have not yet nucleated; iv) 
the nucleation of $\Delta$s softens $P(e)$ over
intermediate energy-densities and stiffens it for high $e$; the
magnitude of the modification increases with $\Delta$ abundances and,
thus, with $(S/A)$.  These results are in agreement with
  \citep{Li_PLB_2018}, who discussed in detail the effect of $\Delta$s
  on the EoS of cold $\beta$-equilibrated matter.

\subsection{Global properties of hot compact stars}
\label{ssec:prop}

Now we turn to the discussion of global properties of hot compact
stars using as input the EoS models presented in the previous
sections. Before discussing our detailed findings for hot stars, let
us recall some general relations observed in older,
$\beta$-equilibrated stars.

\begin{itemize}
  
\item The maximum mass of a compact star is sensitive to the
  interactions in the high-density domain and serves as a useful
  diagnostics of the composition of matter, in particular, nucleation
  of heavy baryons and/or quark matter
  \citep{Weissenborn2012b,Weissenborn2012a,Bonanno2012A&A,Colucci_PRC_2013,Miyatsu_PRC2013,Dalen2014,Gusakov_MNRAS2014,Oertel2015,Fortin_PRC_2016,Fortin2017},

\item The radius of a canonical mass $M_G\simeq 1.44M_{\odot}$  compact star 
  significantly constrain the intermediate-density domain, where 
  the value of the symmetry energy and its slope play an important role
  \citep{Steiner_ApJ_2010,Lattimer_EPJA_2014}. 

\item The tidal deformabilities of compact stars are strongly
  correlated with the radius of the star \citep{Postnikov_PRD_2010}
  and put constraint(s) on the intermediate-density range of the EoS
  \citep{Raaijmakers_2019}.  Their measurement in the GW170817 event
  \citep{Abbott_2017} has already ruled out stiff EoS
  \citep{Most_PRL_2018,Paschalidis_PRD_2018}.
  
\item The compact star moment of inertia depends sensitively to the
  EoS in the intermediate- to low-density regime
  \citep{Ravenhall_ApJ_1994,Lattimer_ApJ_2001,Bejger_MNRAS_2005,Lattimer_ApJ_2005}.
  The difference between the moments of inertia of hypernuclear and
  nucleonic stars and the amount of strangness supported by the
  hypernuclear star are strongly correlated \citep{Fortin_PRD_2020}.

\item The thermal evolution of compact stars is a sensitive probe of
  their interior physics and depends on the EoS mostly via the
  composition of matter in the dense interiors and occurrence of
  various direct Urca processes at high densities. Other uncertainties
  include the composition of the atmosphere and {pairing} gaps of
  various baryonic species that experience attractive
  interactions~\citep{Sedrakian2019EPJA}.  {Conservative models based
    on ``minimal cooling paradigm'' which assume nucleonic stars and
    absence of dUrca processes are so far consistent with the data
    ~\citep{Page_ApJSS_2004,Page_ApJ_2009}. Recent studies of the thermal
    evolution of hypernuclear compact stars with modern EoS
    \citep{Raduta_MNRAS_2018,Raduta_MNRAS_2019,Tolos_ApJ_2018,Grigorian_NPA_2018}
    show that
    agreement with the data can  be achieved in this case as well due to a combination
    of accelerated cooling via various dUrca processes and suppression
    of their rates by superfluidity of heavy baryons. Still, any
    evidence for accelerated cooling cannot be attributed to heavy
    baryonic cores of compact stars, as models which feature nucleonic
    dUrca (see {\it e.g.} \cite{BY15,Wei_MNRAS_2019}) or quark
    matter~\citep{Hess_PRD_2011,deCarvalho_PRC_2015,Sedrakian_EPJA_2016,Wei2020}
    in the centers of compact stars predict accelerated cooling as
    well.  An accelerated cooling can also result from the emission
    of particles beyond the standard model, such as QCD axions or
    axion-like
    particles~\citep{Sedrakian_axion_1,Sedrakian_axion_2,Beznogov2018,Leinson2019JCAP}.
  }
 \end{itemize}

\begin{figure}
  \begin{center}
  \includegraphics[angle=0, width=0.99\columnwidth]{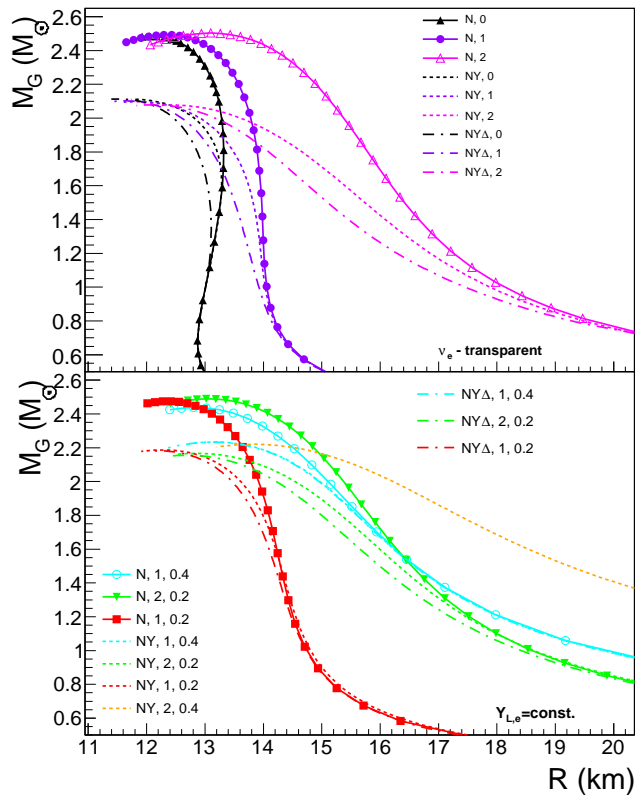}
  \end{center}
  \caption{Gravitational mass $M_G$ versus radius for non-rotating
    spherically-symmetric stars for the DDME2($Y\Delta$) EoS models.
    Top panel: $\beta$-equilibrated, neutrino-transparent stars for
    different values of $S/A$. Bottom panel: Stars with constant
    electron lepton fraction for different values of $S/A$ as
    indicated in the legend in the format $S/A, Y_{L,e}$.  }
  \label{fig:MgR} 
 \end{figure}   

\begin{figure}
  \begin{center}
 \includegraphics[angle=0, width=0.99\columnwidth]{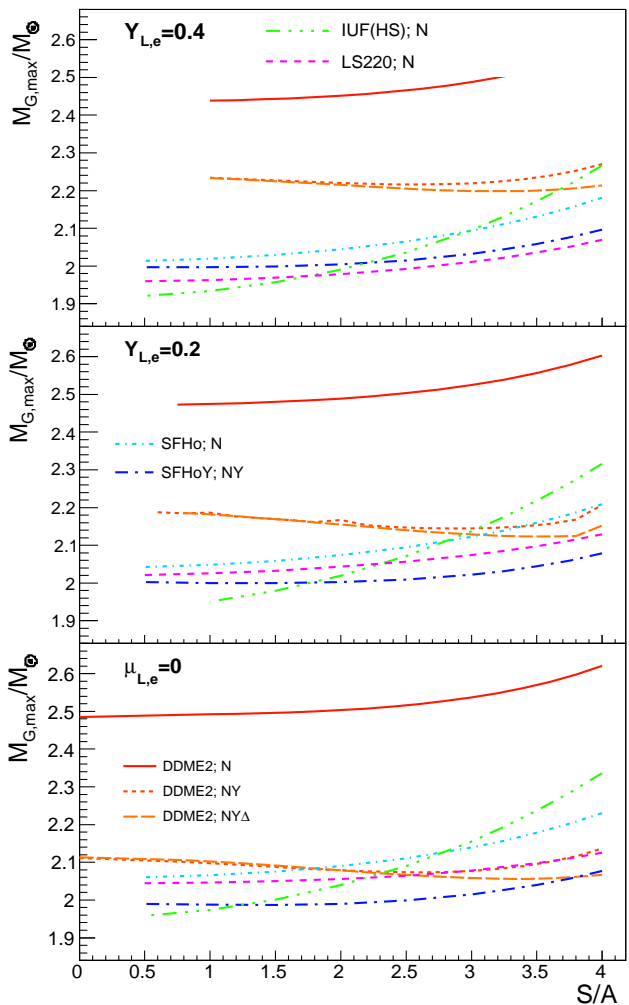}
   \end{center}
  \caption{Maximum gravitational mass $M_{G,{\rm max}}$ versus entropy per baryon
    $S/A$ for non-rotating spherically-symmetric compact stars
    based on the DDME2($Y\Delta$), HS(IUF)~\citep{Fischer2014},
    LS220~\citep{LS_NPA_1991},
    SFHo~\citep{Steiner_2013} and SFHoY~\citep{Fortin_PASA_2018} models
    for $\mu_{L,e}=0$ (bottom) as well as
    $Y_{L,e}=0.2$ (middle) and 0.4 (top panel).
    Results for purely nucleonic
    stars and stars with admixtures of heavy baryons are considered.
  }
  \label{fig:Mgmax_SperA} 
 \end{figure}   

\begin{figure}
  \begin{center}
  \includegraphics[angle=0, width=0.99\columnwidth]{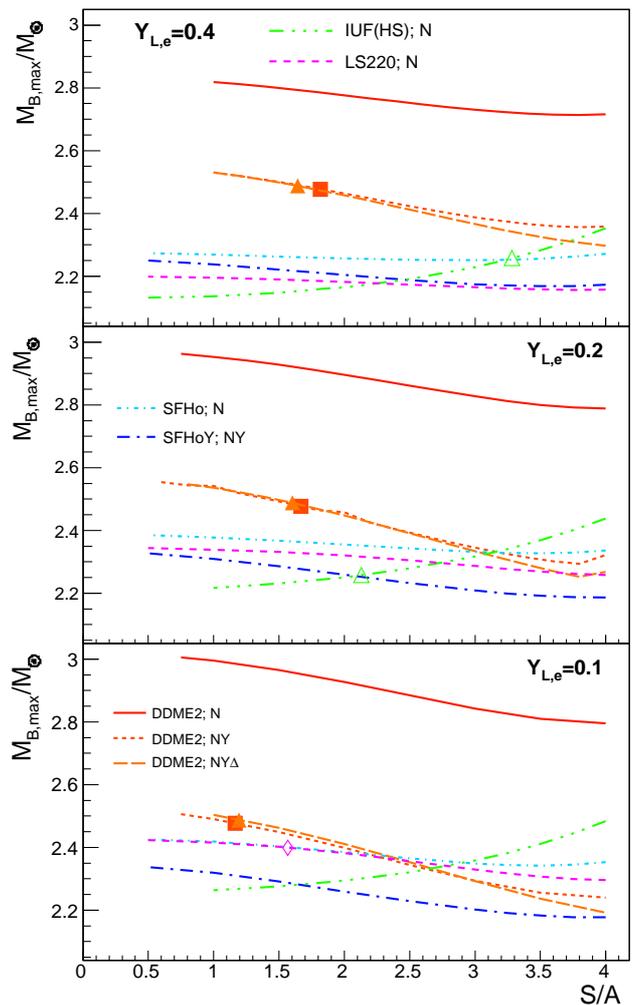}
  \end{center}
  \caption{Maximum baryonic mass $M_{B,{\rm max}}$ versus entropy per baryon $S/A$
    for non-rotating spherically-symmetric {compact stars} {based on }
    the DDME2($Y\Delta$), HS(IUF)~\citep{Fischer2014},
    LS220~\citep{LS_NPA_1991},
    SFHo~\citep{Steiner_2013} and SFHoY~\citep{Fortin_PASA_2018} EoS models
    for $Y_{L,e}=0.1$ (bottom), 0.2 (middle) and 0.4 (top panel).
    Results for purely nucleonic
    stars and stars with admixtures of heavy baryons are considered.
    The value of $S/A$ where the maximum baryonic mass at
    $Y_{L,e}={\rm const.}$ equals the
    maximum baryonic mass of the corresponding cold $\beta$-equilibrated,
    neutrino-transparent star is marked by 
    solid square (DDME2(NY)), solid triangle (DDME2(NY$\Delta$),
    open triangle (IUF(HS)), and diamond (LS220).
  }
  \label{fig:Mbmax_S} 
\end{figure}   

\begin{table*}
  \begin{tabular}{|lc|cc|cc|cc|}
    \hline 
    Model    & $M_{B,{\rm max}}/M_{\odot}$ & $Y_{L,e}$ & instab. domain     & $Y_{L,e}$ & instab. domain & $Y_{L,e}$ & instab. domain \\ \hline
    HS(IUF)  &  2.26                  & 0.4      & $S/A \geq 3.28$     & 0.2      & $S/A \geq 2.13$ &          & \\
    LS 220   &  2.40                  & $~$      & $~$                 &          &                 & 0.1      & $S/A \leq 1.57$ \\
    SFHo     &  2.45                  &          &                     &          &                 &          &  \\
    SFHoY    &  2.36                  &          &                     &          &                 &          & \\
    DDME2    &  3.02                  &          &                     &          &                 &          & \\
    DDME2Y   &  2.48                  & 0.4      & $S/A \leq 1.82$     & 0.2      &  $S/A \leq 1.67$ & 0.1      &  $S/A \leq 1.16$ \\
    DDME2Y$\Delta$ & 2.49             & 0.4      & $S/A \leq 1.64$     & 0.2      &  $S/A \leq 1.60$ & 0.1      &  $S/A \leq 1.19$ \\
    \hline
     \end{tabular}
  \caption{Maximum baryonic masses of cold catalyzed neutrino-transparent
 {compact stars}  {based on }
different EoS models and
    domains of instability with respect to collapse to BH \citep{Bombaci_AA_1996},
    for $Y_{L,e}=0.1,~0.2,~0.4$.}
    \label{tab:stab}
\end{table*}

Fig. \ref{fig:MgR} illustrates the dependence of the star's
gravitational mass on the circumferential radius.
$\beta$-equilibrated stars and stars with constant electron lepton
fraction are considered for different values of entropy per
baryon. The results for baryonic matter composed of $N$, $NY$,
$NY\Delta$ are compared  {assuming} $Y_{L,\mu} = 0$.

Let us start the discussion with the results at zero temperature,
depicted in the top panel of Fig. \ref{fig:MgR}.  Naturally, the
population of additional degrees of freedom such as hyperons or
$\Delta$s modifies both  {compact star} masses and radii.  Nucleation of hyperons
entails a significant reduction of the maximum mass; if in addition
$\Delta$s are accounted for, the maximum mass is only slightly
modified; see Table ~\ref{tab:NS}, too. 

Both the population of hyperons and $\Delta$s reduce the star's
radius. Since the critical density for the onset of $\Delta$s is lower
than that for hyperons, the effect on the radius is visible for the
DDME2Y$\Delta$ parameterisation at lower masses than for the DDME2Y
parameterisation, see Table~\ref{tab:NS} for the respective onset
masses. Since for a canonical mass star  $1.44 M_{\odot}$, hyperons are
present only in a very small amount, the impact on the radius is
small, whereas, within DDME2Y$\Delta$, the reduction is noticeable
(0.16 km), see Table~\ref{tab:NS}.  This finding is in agreement with
the results by
\cite{Spinella-PhD,Li_PLB_2018,Li2020PhRvD,Li2019ApJ,Ribes_2019},
where different underlying nucleonic EoS models have been employed.

Let us now turn to the case of hot {compact} stars.  Top and bottom
panels respectively depict results corresponding to
$\beta$-equilibrated stars and stars with constant (electron) lepton
fraction.  A word of caution is necessary here which concerns, in
particular, the shown radii. For cold {compact stars} it is well known
that radii are sensitive to the crust EoS and the matching of core and
crust~\citep{Fortin_PRC_2016}.  Good experimental constraints on the
{compact star} outer crust composed of stable nuclei combined with a
thermodynamically consistent treatment employing the same interaction
over the whole density range, limit these uncertainties to
$\approx 5\%$ \citep{Fortin_PRC_2016}.  Finite-temperature EoS are
expected to be affected by consistency issues related to the
transition to a clustered matter close to the surface, too. An
additional problem arises since in general the tabulated
finite-temperature EoS contain entries for $T \ge 0.1$ MeV.  For the
EoS studied in this work, the lowest density at which a solution for
the desired $S/A$ can be found in the tables lies in the range
$n_{ll} \approx 10^{-8} - 10^{-7}$ fm$^{-3}$, the exact values
depending on the EoS and the $S/A$ and $Y_{L,e}$ values.  To define
the surface in a coherent way for all models and ($S/A$, $Y_{L,e}$)
conditions, we choose a common $n_{min} \approx 10^{-15}$ fm$^{-3}$
and extrapolate all EoS for $n_{min} \leq n_B < n_{ll}$ with linear
dependencies $\log(n_B)-\log(e)$ and, respectively,
$\log(n_B)-\log(P)$.  This extrapolation together with the arbitrarily
chosen transition density $n_{t} = n_s/2$ between the homogeneous and
inhomogeneous matter lead to some uncertainties in compact star radii
and radius-dependent quantities, {\it e.g.} compactness $C=M_G/R$, the
moment of inertia, quadrupole moment and tidal deformability. Masses
are not affected.  For $S/A \lesssim 2$ the uncertainties on the radii
are smaller than a few percents and those on other quantities even
smaller, see Appendix.  As such this somewhat arbitrary
treatment of the compact star surface affects neither the properties
of hot stars discussed in this section nor the conclusions of Sections
\ref{sec:scalingC} and \ref{sec:scaling}, where universal relations
between different compact star properties are addressed.

The dependence of the maximum gravitational mass $M_{G,{\rm max}}$ on
the star's entropy per baryon is shown in Fig. \ref{fig:Mgmax_SperA}.
Note that the maximum masses in Fig. \ref{fig:Mgmax_SperA} have been
determined at constant total entropy $\left(S/A\right) M_B$, where $M_B$ denotes
the star's baryonic mass following the turning point criterion for a
configuration to be secularly stable; see
\citep{Sorkin_ApJ_1982,Goussard_1998,Marques_PRC_2017} for a detailed
discussion.  The following features are observed in Figs.
\ref{fig:MgR} and \ref{fig:Mgmax_SperA}:
\begin{itemize}
\item[i)] for purely nucleonic stars thermal effects increase the
  gravitational mass and, thus, $M_{G,\rm max}$, whereas stars with an
  admixture of heavy baryons manifest a non-monotonic dependence of
  $M_{G,\rm max}$ on $S/A$. The reason is that as long as thermal
  effects favor nucleation of new species, the maximum mass decreases
  with $S/A$; as soon as all available degrees of freedom are
  populated, we recover the behavior observed for purely nucleonic stars,
  {\it i.e.} $M_{G,\rm max}$ increases with $S/A$.
\item[ii)] 
  for purely nucleonic stars at fixed $S/A$
  $M_{G, \rm max}(\mu_{L,e}=0)> M_{G, \rm max}(Y_{L,e}=0.2)>M_{G, \rm
    max}(Y_{L,e}=0.4)$
with the exception of LS220 at $S/A>3.5$.
  Compact stars with an admixture of heavy baryons most
  frequently show the opposite effect, {\it i.e.} $M_{G, \rm max}$
  increases with $Y_{L,e}$; the reason lies in fact that a lepton rich
  environment with large $Y_{L,e}$ disfavors heavy baryons, such that
  they become less populated and the maximum mass can thus increase
  with $Y_{L,e}$ for given $S/A$; out of the considered cases the only
  exception to this rule is the case of  SFHoY EoS with $S/A\leq 1$ for
  which $M_{G,\rm max}(Y_{L,e}=0.4)<M_{G,\rm max}(Y_{L,e}=0.2)$.
\item[iii)] higher values of $Y_{L,e}$ reduce the star's compactness,
  see Fig. \ref{fig:MgR};
\item[iv)] for a given mass and composition, radii increase with $S/A$,
  {\it i.e.} thermal effects reduce the star's compactness,
  see Fig. \ref{fig:MgR}; 
\item[v)] the magnitude of thermal effects depends on the EoS;
  for DDME2 thermal effects are smaller than those due to $Y_{L,e}$
  and the number of allowed degrees of freedom;
  for LS220 thermal effects dominate over those related to $Y_{L,e}$.
\item[vi)] nucleation of $\Delta$s reduces the gravitational mass,
  including $M_{G,\rm max}$.  Since the effect scales with
  $\Delta$ abundance, it increases with $S/A$ and decreases with
  $Y_{L,e}$.
\end{itemize}

The maximum gravitational mass of isentropic compact stars in
$\beta$-equilibrium is relevant for BH formation in a failed CCSN. By
performing many simulations, \cite{Schneider_2020} have recently shown
that BH formation occurs soon after the PNS's gravitational mass
overcomes $M_{G,\mathit{max}}$ corresponding to its most common
entropy value. The trajectory is thereby essentially determined by the
progenitor compactness, such that the EoS dependence enters mainly via
the behavior of $M_{G,\mathit{max}}$ as function of $S/A$.

The maximum baryon mass is an interesting quantity in the context of
stability against collapse to a black hole during PNS and BNS merger
evolution. In the absence of accretion, $M_B$ is a conserved quantity
during evolution, such that if it exceeds the maximum baryon mass of
the cold $\beta$-equilibrated configuration, the star necessarily
becomes unstable against collapse to a black hole at some point
independently of the mechanism stabilising it temporarily, be it
strong differential
rotation~\citep{Baumgarte1999,Morrison2004,Kastaun2015}, the lepton
rich environment in PNS~\citep{Prakash_1997} or thermal
effects~\citep{Prakash_1997,Kaplan2013}. Following this reasoning,
\cite{Bombaci_AA_1996,Prakash_1997} conjectured that thermally
populated non-nucleonic degrees of freedom lead to the existence of
meta-stable objects, i.e. stars which during the PNS evolution have a
larger maximum baryonic mass than their cold, $\beta$-equilibrated
counterparts and which upon deleptonization and cooling necessarily
collapse to a black hole.

Our results for the maximum baryonic mass, shown as a function of
entropy per baryon $S/A$ for constant (electron) lepton fractions are
plotted in Fig. \ref{fig:Mbmax_S} together with results corresponding
to LS220, HS(IUF), SFHoY and SFHo~\citep{Steiner_2013} EoS models.
The chosen values of the (electron) lepton fraction, $Y_{L,e}=0.1$
(bottom), 0.2 (middle) and 0.4 (top panel), are
relevant for the Kelvin-Helmholtz epoch \citep{Pons_ApJ_1999}.  Again,
the maximum mass has been determined at a fixed total entropy
$\left(S/A\right) M_B$, see \cite{Goussard_1998,Marques_PRC_2017}.
Table \ref{tab:stab} gives, for all models, the values of
$M_{B,{\rm max}}(0, \mu_{L,e}=0)$; also we give, for the considered
$Y_{L,e}$ values, the instability domains defined according to
\cite{Bombaci_AA_1996}.  The values of $S/A$ where
$M_{B,{\rm max}}(S/A,Y_{L,e})=M_{B,{\rm max}}(0,\mu_{L,e}=0)$ are
marked with symbols in Fig. \ref{fig:Mbmax_S}.  DDME2Y($\Delta$)
models show an instability for relatively low $S/A$, whereas the
nucleonic version DDME2 stays stable over the entire range of
considered values for $S/A$. In simulations, during the
Kelvin-Helmholtz phase, $S/A$ stays around 1-2, such that within DDME2
the population of hyperons and/or $\Delta$ might indeed lead to the
formation of a meta-stable object. However, this does not seem to be
possible exclusively in models with non-nucleonic degrees of
freedom. The purely nucleonic HS(IUF) and LS220 show instabilities
too. Although it is not clear whether they will be experienced in the
stellar evolution without performing simulations, it cannot be excluded
that meta-stable nucleonic stars could form.  These results show the
importance of the nuclear interaction in understanding the properties
of the high-density EoS.

\begin{table*}
  \begin{tabular}{|l|cccccc|}
    \hline 
    Thermo. cond. & $p_0$ & $p_1$ & $p_2$  & $p_3$ & $p_4$ & Refs. \\ \hline
    &  $c_0$ & $c_1$ & $c_2$  & $c_3$ & $c_4$ &  \\
    $T=0$, $\beta$-eq. & 0.244 & 0.638 & 0
     & 0 & 3.202 &\citep{Breu_MNRAS_2016} \\
    $S/A=2$, $Y_{L,e}=0.2$ & $5.965 \cdot 10^{-2}$ & 2.35082 &
    -6.7077 &  10.6489 & 0 & this work \\
 \hline 
     &  & $a_1$ & $a_2$ & $a_3$  & $a_4$ &  \\
     $T=0$, $\beta$-eq. & & $8.134 \cdot 10^{-1}$ & $2.101 \cdot 10^{-1}$ &
    $3.175 \cdot 10^{-3}$ &  $-2.717 \cdot 10^{-4}$& \citep{Breu_MNRAS_2016} \\
    $S/A=2$, $Y_{L,e}=0.2$ & &$9.447 \cdot 10^{-1}$ & $1.815 \cdot 10^{-1}$ &
    $-4.049 \cdot 10^{-3}$ &  $4.339 \cdot 10^{-5}$ & this work \\
  \hline
     & &  $b_1$ & $b_2$ & $b_3$  & &  \\
    $T=0$, $\beta$-eq. & & $3.71 \cdot 10^{-1}$ & $-3.91\cdot 10^{-2}$ &
    $ 1.056 \cdot 10^{-3}$ & & \citep{Maselli_PRD_2013} \\
    $S/A=2$, $Y_{L,e}=0.2$ & & $3.632 \cdot 10^{-1}$ & $-4.216\cdot 10^{-2}$
    & $1.288\cdot 10^{-3}$  & & this work \\
  \hline
  & $e_0$ & $e_1$ & $e_2$ & $e_3$ & & \\
  $T=0$, $\beta$-eq. & -2.7157 & 0.7017 & 0.1611 & $-6.4977 \cdot 10^{-3}$ & & this work \\
  $S/A=2$, $Y_{L,e}=0.2$ & -1.8410 & 0.5829 & 0.1081 & $-4.3085 \cdot 10^{-3}$ & & this work \\
  \hline
  & & $d_1$ & $d_2$ & & &  \\
  $T=0$, $\beta$-eq. & & 0.6213 & 0.1941 && &\citep{Breu_MNRAS_2016} \\
  $S/A=2$, $Y_{L,e}=0.2$ & & 0.4733  & 0.9232 & & & this work \\
  \hline   
  \end{tabular}
  \caption{Fitting parameters entering Eqs. 
    \eqref{eq:ItildeC}-\eqref{eq:BE} under different indicated
    thermodynamic conditions.
  }
  \label{tab:params_fit}
\end{table*}

\section{Universal relations}

Although NS properties depend sensitively on the EoS, several
``universal" relations have been found between global quantities.  The
term ``universality" refers here to very weak dependence on the EoS
which holds well for cold $\beta$-equilibrated stars.  Although so far
the reason for this universal behavior is not well understood, it may
be exploited to constrain quantities difficult to access
observationally, eliminate the uncertainties related to the EoS
in the analysis of the data, or break degeneracies between 
integral quantities, ({\it e.g.}, the quadrupole moment and the neutron-star
spins in binary in-spiral waveforms).

In this section, we shall investigate to what extent this universality
remains valid for hot and lepton rich stars, allowing for
various particle degrees of freedom in the EoS.
Section \ref{sec:scalingC} will address relations between the
normalised moment of inertia, quadrupole moment, tidal deformability
and binding energy and the star's compactness, and Section
\ref{sec:scaling} will address the $I$-${\rm Love}$-$Q$ relations.
Binding energies and tidal deformabilities will be calculated for
non-rotating spherically-symmetric stars;
moments of inertia and quadrupole deformations will be calculated
for rigidly and slowly rotating stars.

\begin{figure*}
  \begin{center}
   \includegraphics[angle=0, width=0.45\textwidth]{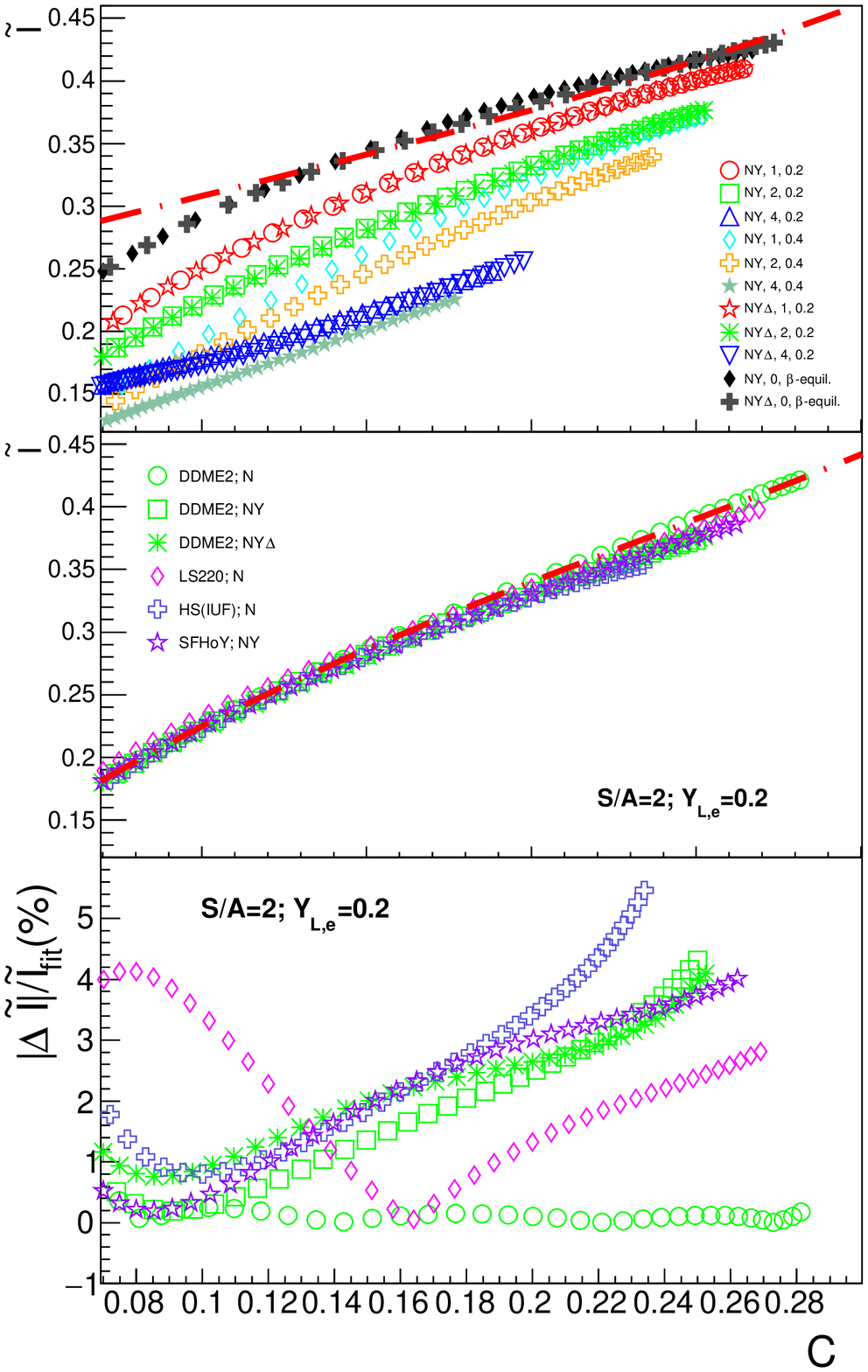}
   \includegraphics[angle=0, width=0.45\textwidth]{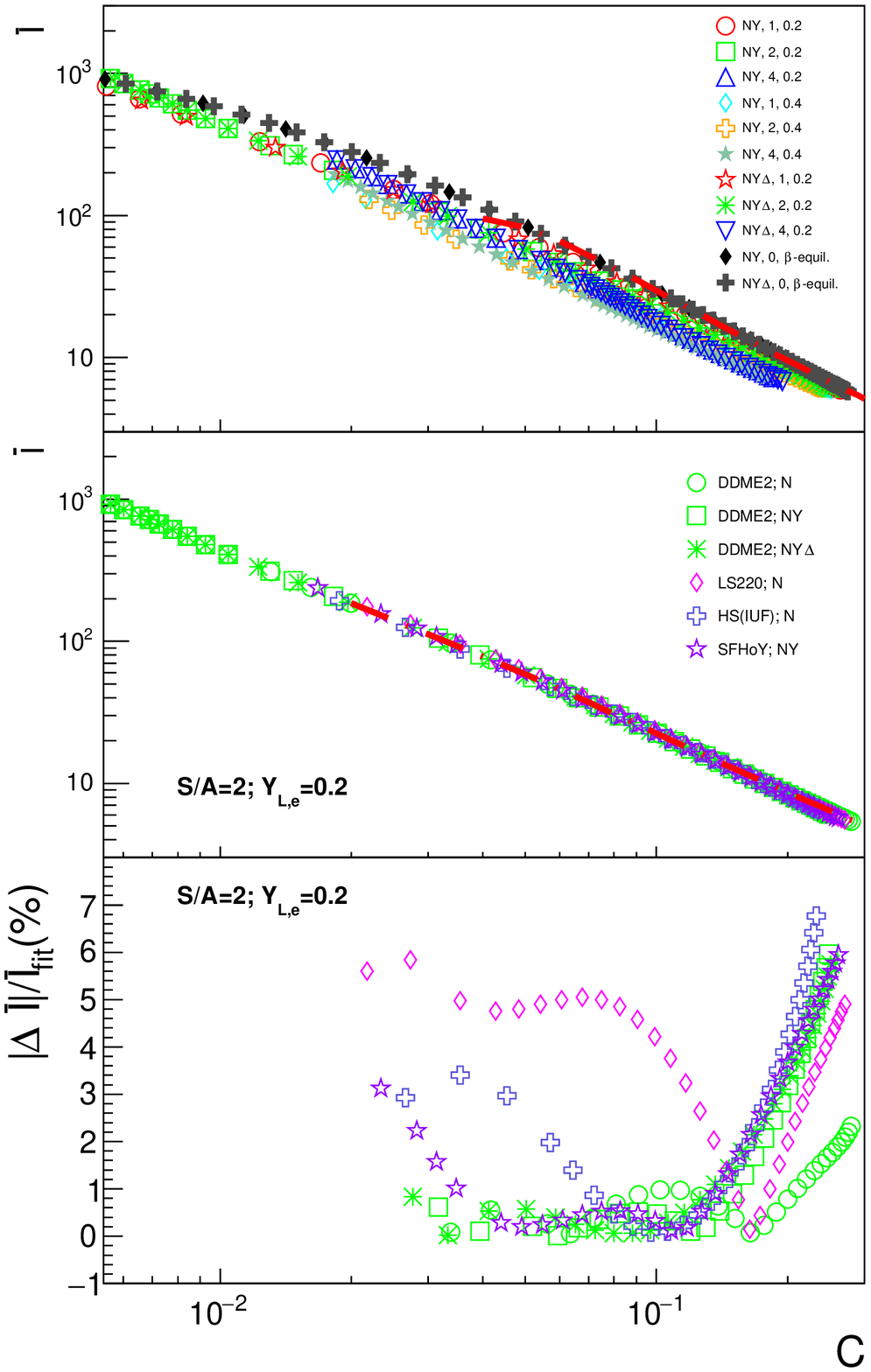}
  \end{center}
  \caption{Normalised moments of inertia
    $\tilde I =I/\left(M_G R^2 \right)$ (left panels) and
    $\bar I=I/M_G^3$ (right panels) as function of compactness
    $C=M_G/R$.  Top panels: results corresponding to different
    thermodynamic conditions and matter compositions, as
     {obtained from the} DDME2(Y$\Delta$) model.
    At finite temperatures the thermodynamic conditions are indicated
    in terms of constant $S/A$ and $Y_{L,e}$; the label ``0,
    $\beta$-equil." corresponds to cold catalyzed neutrino-transparent
    matter.  Middle panels: results corresponding to ($S/A=2$,
    $Y_{L,e}=0.2$) for different matter compositions and EoS models.
    The long dot-dashed curves correspond to the fits in
    Eqs. (\ref{eq:ItildeC}) and (\ref{eq:Ibar}), respectively, with
    parameters from Table~\ref{tab:params_fit}.  Bottom panels:
    relative residual errors with respect to the fits in
    Eqs. (\ref{eq:ItildeC}) (left panel) and (\ref{eq:Ibar}) (right
    panel) for the cases considered in the middle panels.}
  \label{fig:normI_C}  
 \end{figure*}   
\subsection{Dependence on the compactness}
\label{sec:scalingC}

For cold, $\beta$-equilibrated stars in the slow-rotation
approximation, several authors have established relations between the
compactness of a star $C=M_G/R$ and
normalised moment of inertia, quadrupole moment, tidal
deformability and binding energy,
which show universal character, {\it i.e.} are almost
EoS independent.

First, by considering different EoS for cold $\beta$-equilibrated
neutron star matter
\cite{Ravenhall_ApJ_1994} noted that, except for very low
mass stars, the normalised moment of inertia $\tilde I=I/\left(M_GR^2 \right)$
behaves as a universal function of the star's mass and radius,
\begin{equation}
  \tilde I \approx 0.21 \frac{1}{1-2M_G/R}.
  \end{equation}
Later, \cite{Lattimer_ApJ_2005} proved that $\tilde I$
can be expressed as a polynomial in compactness:
\begin{equation}
  \tilde I=c_0+c_1 C+ c_2 C^2 + c_3 C^3 + c_4 C^4.
  \label{eq:ItildeC}
\end{equation}
The issue was recently reconsidered by \cite{Breu_MNRAS_2016} who
showed that the dispersion between different EoS in
Eq.~(\ref{eq:ItildeC}) is reduced if only models which fulfill the
$2M_{\odot}$ maximum mass constraint are considered.   {Furthermore,
  they} found another universal relation, relating alternatively
normalised moment of inertia $\bar I=I/M_G^3$ [note the different
normalisation with respect to Eq.~(\ref{eq:ItildeC})] to compactness
\begin{equation}
  \bar I= a_1 C^{-1}+a_2 C^{-2}+ a_3  C^{-3}+ a_4 C^{-4}.
  \label{eq:Ibar}
\end{equation}
\cite{Maselli_PRD_2013} introduced a universal expression relating
compactness to the normalised tidal deformability
$\bar \lambda=\lambda/M_{G}^5$
\begin{equation}
  C=b_1 +b_2\ln{\bar \lambda}
  + b_3 \left( \ln{\bar \lambda} \right)^2.
  \label{eq:CLove}
\end{equation}

 {Earlier,} by considering a collection of different EoS models,
\cite{Yagi_PRD_2013} have shown that the scaled quadrupole moment
$\bar Q=Q M_G/J^2$, where $J$ stands  {for} the angular momentum,
as function of compactness is only weakly dependent on the EoS. Our
results suggest that $\bar Q$ can be expressed as a polynomial of
$C^{-1}$
 \begin{equation}
   \bar Q=e_0+e_1 C^{-1}+e_2 C^{-2}+e_3 C^{-3} .
   \label{eq:QC}
   \end{equation}
 {which is analogous to Eq \eqref{eq:Ibar} for $\bar I$.}
   The neutron star binding energy, defined as the difference between
   baryonic and gravitational masses $E_B=M_B-M_G$, shows little
   sensitivity on the underlying EoS model if normalised by the
   gravitational mass \citep{Lattimer_ApJ_2001,Breu_MNRAS_2016}.
   According to \cite{Lattimer_ApJ_2001} it behaves  {as}
\begin{equation}
  \frac{E_B}{M_G}=\frac{d_1  C}{1-d_2 C}~.
  \label{eq:BE}
\end{equation}
Again, as in the case of $\tilde I(C)$, limiting the considered EoS
models to those  {which are consistent
  with} the $2M_{\odot}$ mass constraint improves the quality of the
fit~\citep{Breu_MNRAS_2016}.
The values of the different fitting parameters, $a_i$, $b_i$, $c_i$, $d_i$, $e_i$,
entering Eqs.~\eqref{eq:ItildeC}-\eqref{eq:BE} are provided in
Table~\ref{tab:params_fit}.

\begin{figure}
  \begin{center}
    \includegraphics[angle=0, width=0.45\textwidth]{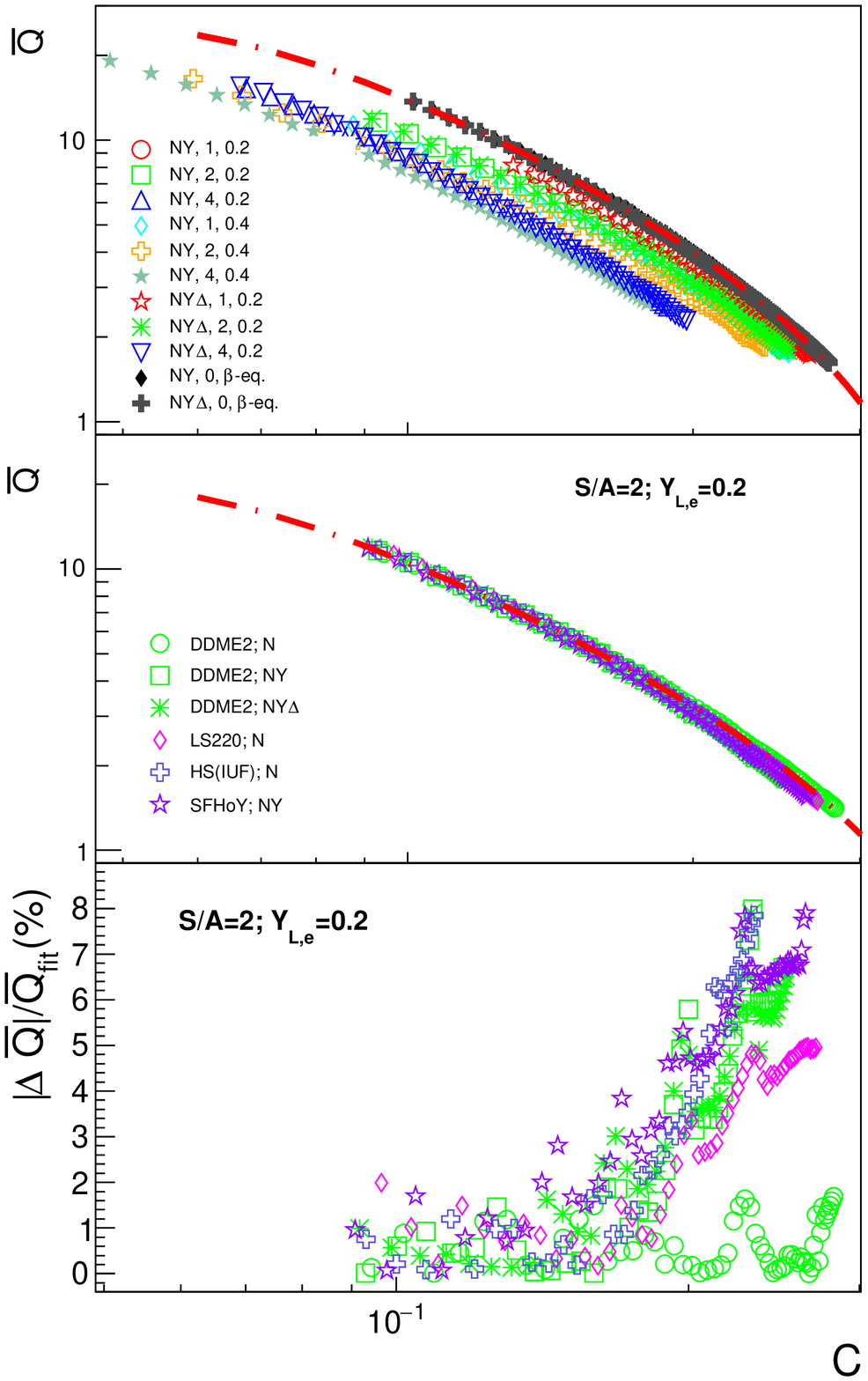}
  \end{center}
  \caption{Normalised quadrupole moment $\bar Q$ of slowly and
    rigidly rotating stars as function of compactness, $C$.
    Top panel: results corresponding to different thermodynamic conditions
    (mentioned in the legend)
    and matter compositions  {derived from the
    DDME2(Y$\Delta$) model}.
    At finite temperatures the thermodynamic conditions are specified
    in terms of constant $S/A$ and $Y_{L,e}$;
    the label ``0, $\beta$-equil." corresponds to cold catalyzed
    neutrino-transparent matter.
    Middle panel: results corresponding to ($S/A=2$, $Y_{L,e}=0.2$) for
    different EoS models and  {for} different matter
    compositions.  The red dot-dashed curves correspond to fits in
    Eq. (\ref{eq:QC}) with parameters values from Table
    \ref{tab:params_fit}.  Bottom panel: relative residual errors with
    respect to the fit in Eq. (\ref{eq:QC}) for the cases considered
    in the middle panels.  }
  \label{fig:Q_C}
\end{figure}   
 
 \begin{figure}
  \begin{center}
   \includegraphics[angle=0, width=0.45\textwidth]{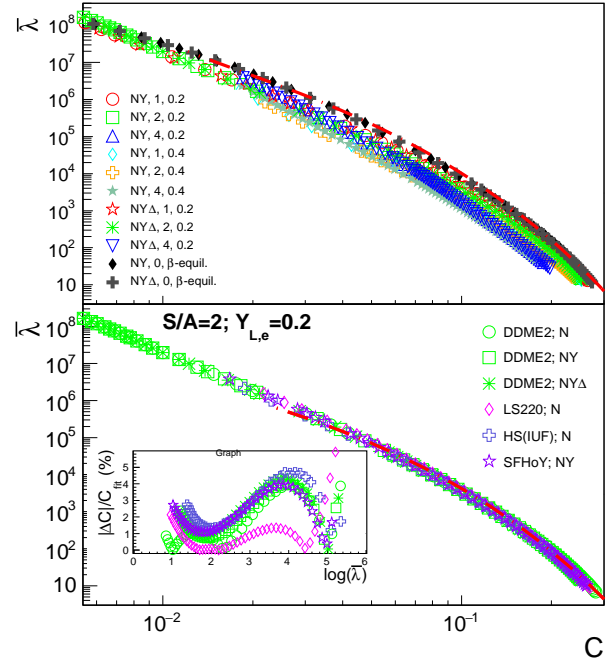}
  \end{center}
  \caption{ Normalised tidal deformability $\bar \lambda$ of
    non-rotating spherically-symmetric stars as function of
    compactness $C$.  Top panels: results corresponding to different
    thermodynamic conditions (mentioned in the legend) and matter
    compositions, as  {obtained}  {from the DDME2(Y$\Delta$)
      model}.  At finite temperatures the thermodynamic conditions are
    specified in terms of constant $S/A$ and $Y_{L,e}$; the label ``0,
    $\beta$-equil." corresponds to cold catalyzed
    neutrino-transparent matter.  Bottom panels: results
    corresponding to ($S/A=2$, $Y_{L,e}=0.2$) for different EoS models
     {for} different matter compositions.  The red dot-dashed
    curves correspond to fit in Eq. (\ref{eq:CLove}) with parameters
    values from Table \ref{tab:params_fit}.  Relative residual errors
    with respect to the the fit in Eq. (\ref{eq:CLove}) are
     {shown} in the inset.  }
  \label{fig:L_C}  
 \end{figure}   

\begin{figure}
  \begin{center}
    \includegraphics[angle=0, width=0.99\linewidth]{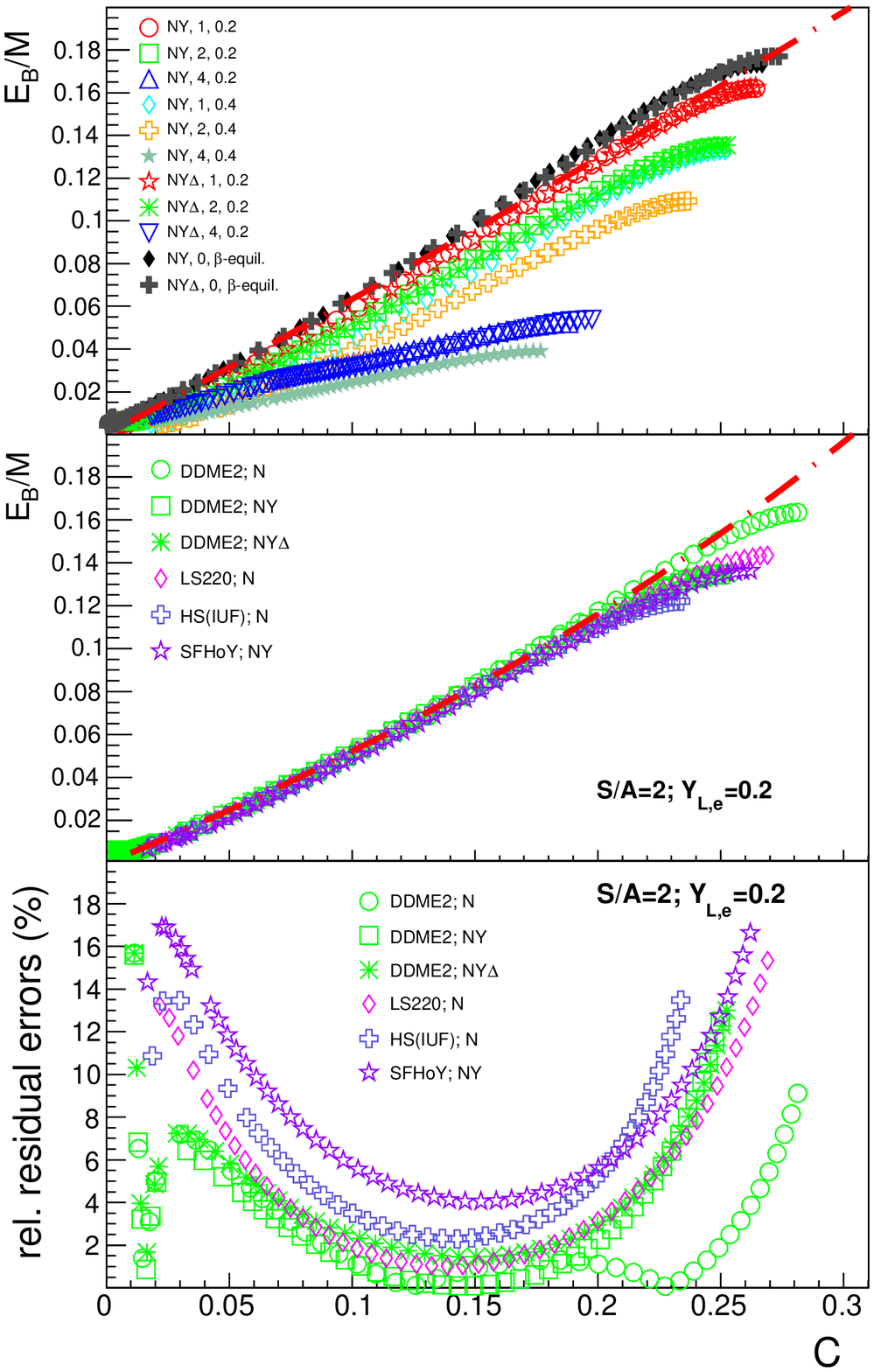}
  \end{center}
  \caption{ Binding energy per unit of gravitational mass $E_B/M_G$ of
    non-rotating spherically-symmetric stars as a function of
    compactness $C$.  Top panel: results corresponding to different
    thermodynamic conditions (indicated in the legend) and matter
    compositions, as {obtained from the} DDME2(Y$\Delta$) {model}.  At
    finite temperatures the thermodynamic conditions are specified in
    terms of constant $S/A$ and $Y_{L,e}$; the label ``0,
    $\beta$-equil." corresponds to cold catalyzed neutrino-transparent
    matter.  Middle panel: results corresponding to ($S/A=2$,
    $Y_{L,e}=0.2$) for different EoS models and {for} different matter
    compositions.  The red dot-dashed curves correspond to fits in
    Eq. (\ref{eq:BE}) with parameters values from Table
    \ref{tab:params_fit}.  Bottom panel: relative residual errors with
    respect to the fit for the cases considered in the middle panel.
  }
    \label{fig:B_C}  
\end{figure}   

Let us now turn to the discussion of universality at nonzero entropy.
The top panels of Figs. \ref{fig:normI_C}, \ref{fig:Q_C},
\ref{fig:L_C} and \ref{fig:B_C} illustrate, respectively, the behavior
of $\tilde I$, $\bar I$, $\bar Q$, $\bar \lambda $, and $E_B/M_G$ as a
function of compactness for different combinations of constant $S/A$
and $Y_{L,e}$, as  {obtained from the}
DDME2(Y$\Delta$)  {model}.  The moment of inertia is thereby
calculated to leading order in the slow, rigid rotation
approximation~\citep{Hartle_ApJ_1967} and the tidal deformability
$\lambda$ is computed following \cite{Hinderer_ApJ_2008} and
\cite{Hinderer_PRD_2010}.  The quadrupole moment is 
computed for a rotation frequency of 100 Hz using the public domain
code Lorene\footnote{\url{https://lorene.obspm.fr}}
\citep{Lorene}. For comparison, for cold, $\beta$-equilibrated matter
the results  {obtained} from  {the} DDME2(Y$\Delta$)  {model}
as well as the predictions  {of}
Eqs.~\eqref{eq:ItildeC}-\eqref{eq:BE} are shown, see
Table~\ref{tab:params_fit} for the parameter values.

For all these quantities, a significant scatter of the results
for different values of $S/A$ and $Y_{L,e}$ is observed.
The deviation from the results corresponding to cold catalyzed matter increases
with $S/A$ and/or $Y_{L,e}$.
This indicates that universality does not hold when stars with
different entropies and lepton contents are compared.
For $\tilde I$ and $\bar I$ this conclusion has recently been reached by
\cite{Lenka_JPG_2019}.   {However, one needs in fact to compare
  quantities under identical thermodynamic conditions.}   {To
  illustrate this point, we} display in the middle panels of
Figs. \ref{fig:normI_C}, \ref{fig:Q_C} and \ref{fig:B_C} and bottom
panel of Fig. \ref{fig:L_C} the results for different EoS models and
compositions,  {for} the case ($S/A=2$,
$Y_{L,e}=0.2$).  In addition to matter made of $N$, $NY$ or $NY\Delta$
based on the DDME2(Y$\Delta$) models, we show results corresponding to
the purely nucleonic LS220 and HS(IUF) EoS models
\citep{LS_NPA_1991,Fischer2014} as well as the hyperonic SFHoY EoS
\citep{Fortin_PASA_2018}.  The agreement between the different EoS
models is very good, except for $\tilde I$, $\bar Q$, and $E_B$ at
large values of $C \gtrsim 0.2$, where some deviations can be seen.
This confirms that indeed universality holds well for all these
relations if the same thermodynamic conditions are considered.

To quantify this universality, we have performed fits to the nucleonic
DDME2 results following Eqs.~(\ref{eq:ItildeC})-(\ref{eq:BE}),
indicated in the figures by the green open circles.  The corresponding
parameter values are listed in Table~\ref{tab:params_fit}. Overall,
the good description of the results indicates that the functional
relations proposed for cold $\beta$-equilibrated matter hold for hot
matter with trapped neutrinos, too. Only for $\tilde I(C)$, the data
are better described by a third-order polynomial, see
Table~\ref{tab:params_fit}.  For $\tilde I(C)$, $\bar I(C)$, $\bar Q$
and $C(\bar \lambda)$ the deviations from the fits are of the order of
a few percents, as illustrated in the bottom panels of
Figs.~\ref{fig:normI_C} and \ref{fig:Q_C} and, respectively, the inset
in the bottom panel of Fig. \ref{fig:L_C}, where the relative residual
errors with respect to the fit functions are depicted.  We conclude
that for these quantities universality holds with a precision only
slightly inferior to that for the cold $\beta$-equilibrated case.  For
$E_B/M_G(C)$ the deviations from the fit are larger, up to $20\%$, see
the bottom panel of Fig.~\ref{fig:B_C}.

\begin{figure*}
\begin{center}
\includegraphics[angle=0, width=0.45\textwidth]{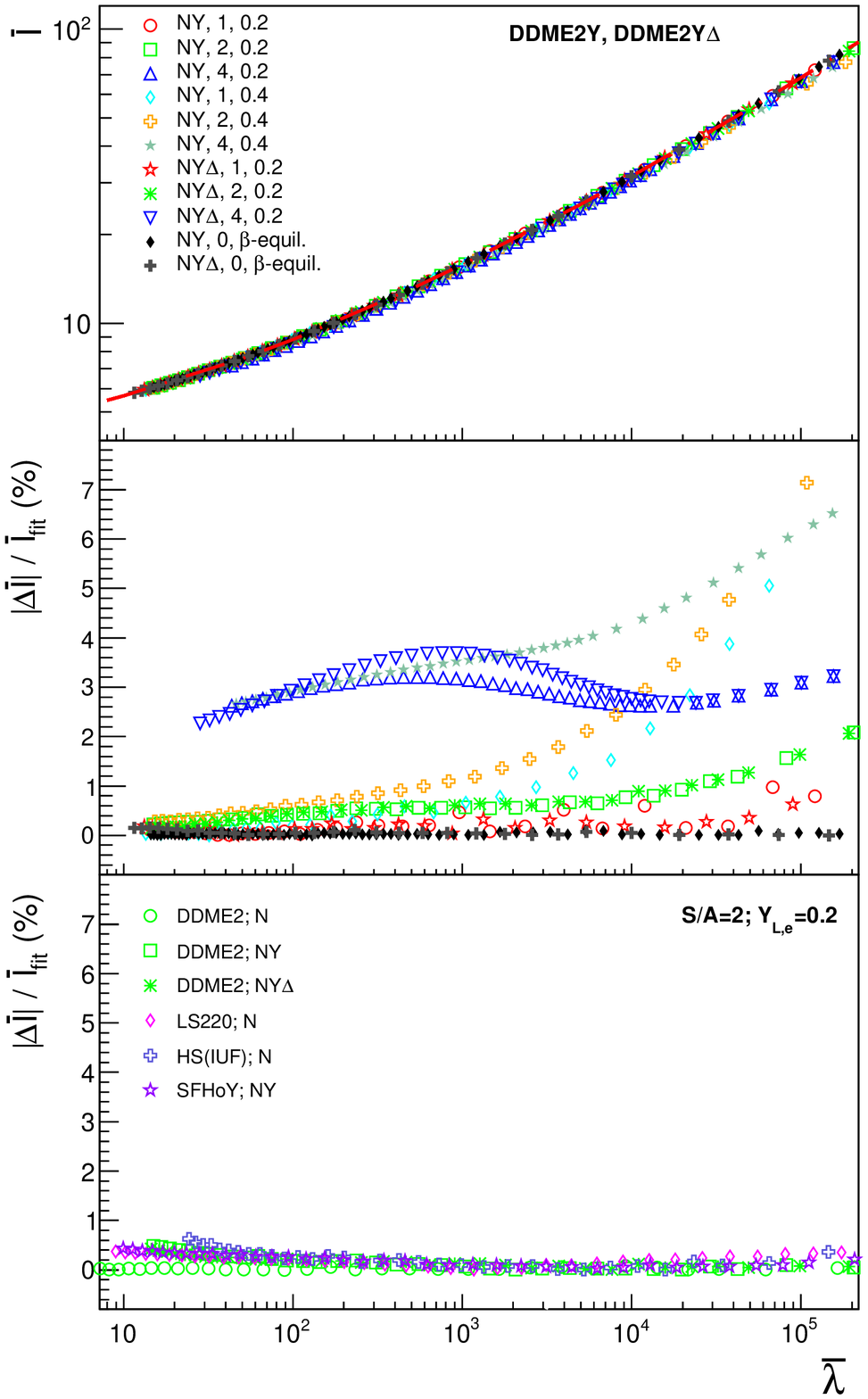}
\includegraphics[angle=0, width=0.45\textwidth]{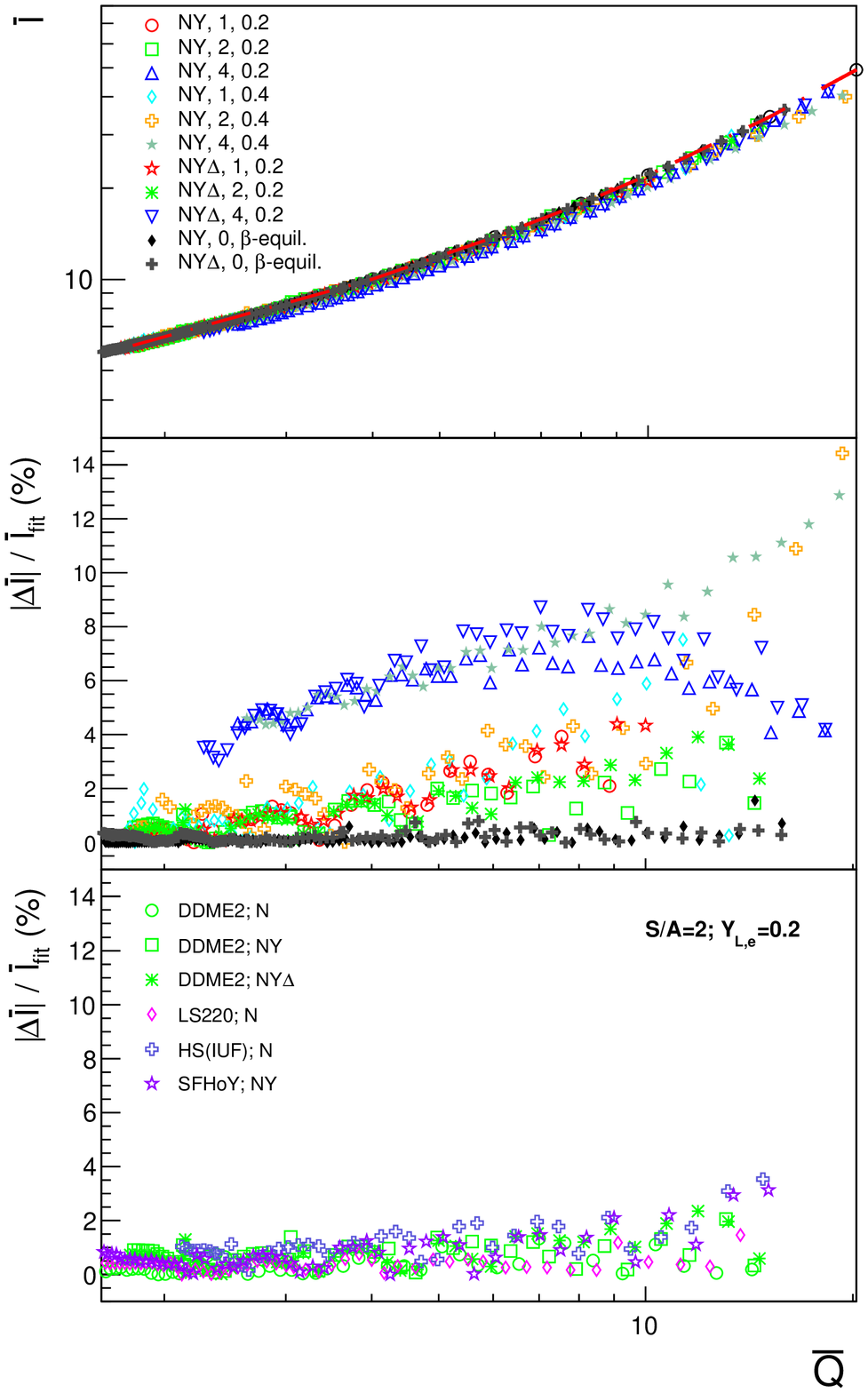}
\end{center}
\caption{Left: $\bar I$ vs. $\bar \lambda$ (top) and relative residual
  errors $|\Delta \bar I|/\bar I_{fit }(\bar \lambda)$ with respect to
  the fit by eq. (\ref{eq:YYfit}) (middle, bottom).  Right: $\bar I$
  vs. $\bar Q$ (top) and relative residual errors
  $|\Delta \bar I|/\bar I_{fit }(\bar Q)$ with respect to the fit by
  eq. (\ref{eq:YYfit}) (middle, bottom).  Top: results for different
  thermodynamic conditions and matter compositions (indicated in the
  legend), as   {obtained from the}
  DDME2(Y$\Delta$)  {model}. The dot-dashed red line
  indicates the fit to the cold $\beta$-equilibrated results, see
  Table~\ref{tab:params_ILQ}. Middle: relative error with respect to
  the fit at zero temperature. Bottom panels: results corresponding to
  $S/A=2$ and $Y_{L,e}=0.2$. The relative error with respect to
  refitted values under these thermodynamic conditions, see
  Table~\ref{tab:params_ILQ}, is shown for different matter
  compositions and EoS models.  }
\label{fig:I-Love-Q}
\end{figure*}
\subsection{$I$-Love-$Q$ universal relations}
\label{sec:scaling}
\begin{figure*}
  \begin{center}
    \includegraphics[angle=0, width=0.45\textwidth]{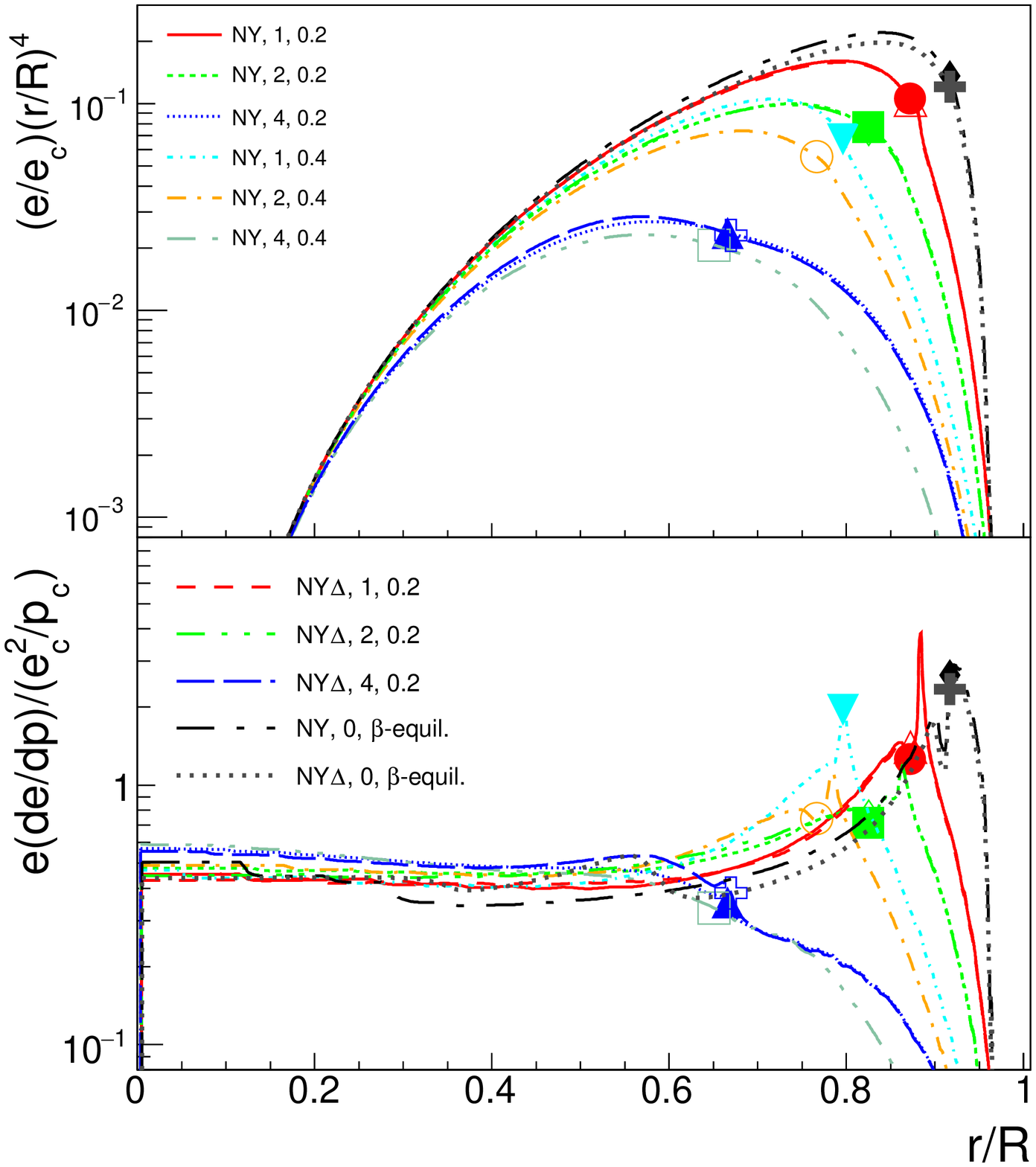}\hfill
    \includegraphics[angle=0, width=0.45\textwidth]{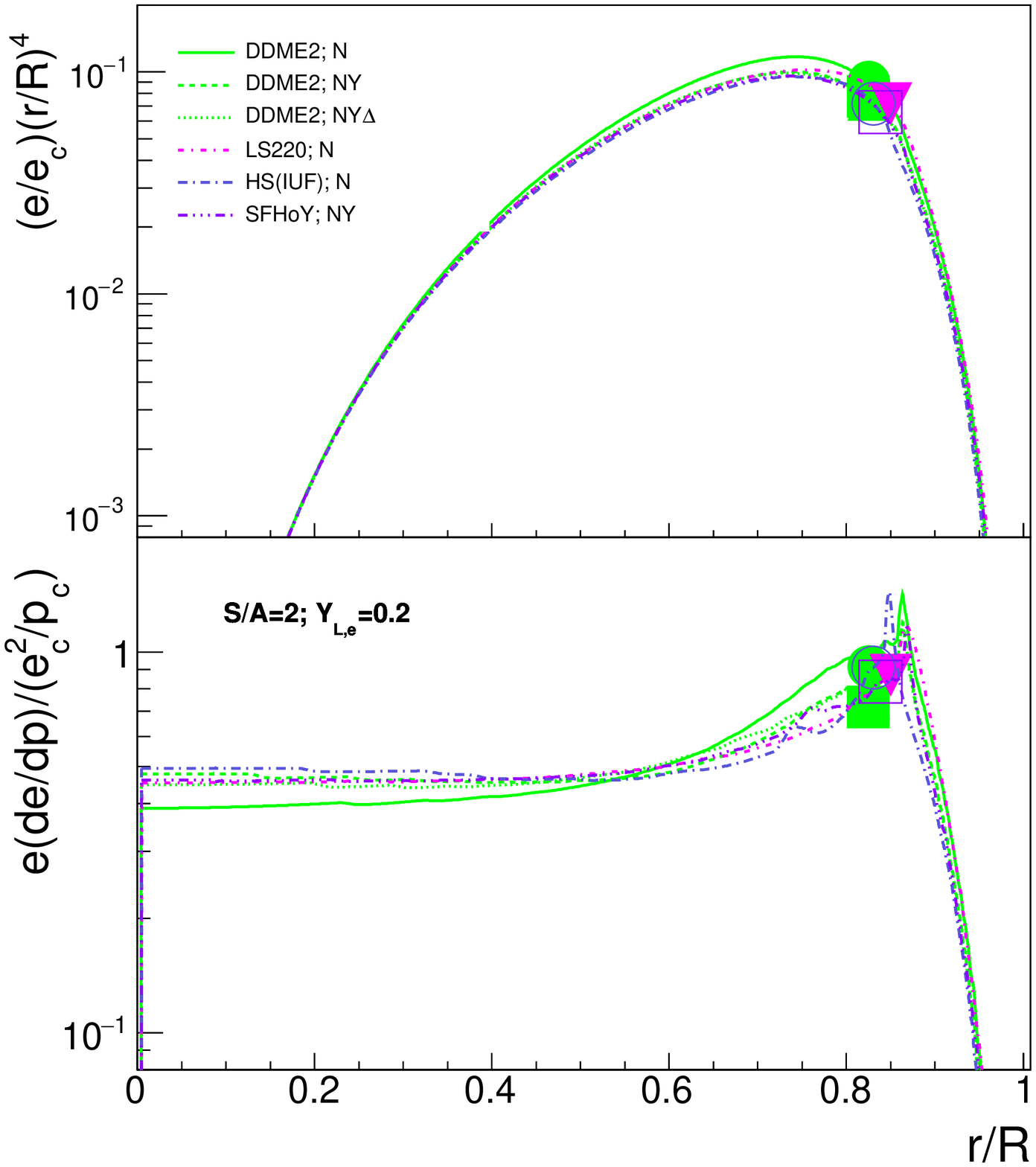}
  \end{center}
  \caption{Normalised radial profiles of
    $\left(e/e_c \right) \left( r/R \right)^4$ (top panels) and
    $e \left( de/dp \right) \left( e_c^2/p_c \right)$ (bottom panels)
    for stars with compactness $C=0.17$, as predicted by various EoS
    models.  Left panels illustrate results corresponding to
    DDME2(Y$\Delta$)  {model} for various thermodynamic conditions
    specified in the legend in terms of $(S/A,Y_{L,e})$; the label
    ``0, $\beta$-equil." corresponds to cold catalyzed
    neutrino-transparent matter.  Right panels illustrate results
    corresponding to different EoS models and matter compositions
    ($N$, $NY$, $NY\Delta$) for $(S/A=2, Y_{L,e}=0.2)$.
  The crust-core transition is indicated, in each case, by a symbol.}
  \label{fig:radprofiles}  
 \end{figure*}   

\cite{Yagi_Science_2013,Yagi_PRD_2013} identified universal relations
among the pairs of quantities $\bar I$ - $\bar Q$, $\bar I$ -
$\bar \lambda$ and $\bar Q$ - $\bar \lambda$.  Numerically, these
relations can be cast in a polynomial form on a log-log scale,
\begin{equation}
  \ln Y_i= a_i + b_i \ln X_i + c_i \left( \ln X_i \right)^2
  + d_i \left( \ln X_i \right)^3 + e_i \left( \ln X_i \right)^4 ,
  \label{eq:YYfit}
\end{equation}
with the pairs $(Y_i,X_i)$ corresponding to
$(\bar I,\bar\lambda);(\bar I, \bar Q);(\bar Q,\bar \lambda)$.
Originally established in the slow rotation limit, these relations
remain EoS independent for fast rotating stars with rotation frequency
dependent fit parameters~\citep{Doneva_2013}.

\begin{table*}
  \begin{tabular}{lllrrrrrc}
    \hline 
    $Y$ & $X$ & Thermo. cond. & $a$ & $b$ & $c$ & $d$ & $e$  & Refs. \\
    \hline
    $\bar I$ & $\bar \lambda$ & $T=0$, $\beta$-eq. & 1.47 & $7.19\times 10^{-2}$
    & $2.00\times 10^{-2}$ &
    $-5.00 \times 10^{-4}$ & $2.39 \times 10^{-6}$ & this work \\
    $\bar I$ & $\bar Q$ & $T=0$, $\beta$-eq. & 1.50 & $4.66\times 10^{-1}$ &
    $6.11\times 10^{-2} $ & $1.30 \times 10^{-2}$ & $1.20 \times 10^{-3}$
    &  this work  \\
    $\bar I$ & $\bar \lambda$ &$S/A=2$, $Y_{L,e}=0.2$ & 1.49 & $6.55 \times 10^{-2}$ &
    $2.06 \times 10^{-2}$ & $-4.47 \times 10^{-4}$ & $-2.96 \times 10^{-6}$ & this work \\
    $\bar I$ & $\bar Q$ & $S/A=2$, $Y_{L,e}=0.2$ & 1.53 & $3.81 \times 10^{-1}$ &
    $1.36 \times 10^{-1}$ & $-1.21 \times 10^{-2}$ & $3.45 \times 10^{-3}$ & this work \\
 \hline
  \end{tabular}
  \caption{Fitting parameters of eq. (\ref{eq:YYfit})
    for different thermodynamic conditions.
  }
  \label{tab:params_ILQ}
\end{table*}

Let us now consider $I-$Love$-Q$ universality at nonzero
temperature. We will focus here on $\bar I-\bar \lambda$ and
$\bar I-\bar Q$.  If universality holds for these two pairs of
quantities, it is very likely that it will hold for the third pair as
well.  In Fig.~\ref{fig:I-Love-Q} we plot $\bar I$ as function of
$\bar \lambda$ (left) and $\bar Q$ (right). Again, in the top panels,
we compare results for different values of $S/A$ and $Y_{L,e}$
employing the DDME2(Y$\Delta$) {models} with those for cold
$\beta$-equilibrated stars. The dot-dashed line indicates the result
of Eq.~(\ref{eq:YYfit}) with fitting parameters obtained for cold
$\beta$-equilibrated stars {according to the} DDME2Y {model}, see
Table~\ref{tab:params_ILQ}.  As can be seen from the middle panels,
the residual errors become an order of magnitude larger than those
obtained when considering only cold $\beta$-equilibrated stars
\citep{Yagi_PRD_2013}. This deviation of the $I-$Love$-Q$-relations
from universality due to thermal effects have already been noted by
\cite{Martinon_PRD_2014,Marques_PRC_2017}.

The discussion in Sec. \ref{sec:scalingC} suggests that universality
at nonzero entropy can be recovered  {under} identical
thermodynamic conditions. The bottom panels of Fig.~\ref{fig:I-Love-Q}
show the relative error of the results with respect to
Eq.~(\ref{eq:YYfit}) with refitted parameters at $S/A = 2$ and
$Y_{L,e} = 0.2$, see Table~\ref{tab:params_ILQ}. Results corresponding
to DDME2(Y$\Delta$) model are confronted with those of LS220
\citep{LS_NPA_1991}, HS(IUF) \citep{Fischer2014} and SFHoY
\citep{Fortin_PASA_2018}. The quality of the fit by
Eq. (\ref{eq:YYfit}) is considerably improved reaching the
 {accuracy}  {of} zero temperature, $\beta$-equilibrated
 {case}. Thus, universality again holds under the same
thermodynamic conditions.

This phenomenologically observed universality is not yet
understood. Analytical solutions in the Newtonian
limit~\citep{Yagi_PRD_2013} {as well as those obtained using an}
expansion around this limit~\citep{Jiang_2020} corroborate the EoS
independence, but without {providing a definitive insight into their
  origin.}  \cite{Yagi_PRD_2013} advanced two possible
explanations. First, these dimensionless quantities mainly depend on
the outermost shells of the core and on the crust, where, by
construction, "realistic" EoS, {\it i.e.} nuclear EoS with
  parameters fitted to nuclear data and/or astrophysical
  observations, agree with each other. As in
  \citep{Yagi_PRD_2013} the term "realistic" is meant here  to
  distinguish nuclear EoS from polytropic EoS.  If this was the case,
finite-temperature EoS should necessarily violate universality as
their low-density behaviors differ from each other, and from that of
cold $\beta$-equilibrated matter, see the discussion in
Sec.~\ref{sec:NS}. Second, \cite{Yagi_PRD_2013} suggested that this
universality could be a reminiscence of no-hair theorems as the
neutron star's compactness approaches the black hole limit. In this
case, too, since hot stars are less compact, see Sec.~\ref{sec:NS},
universality should be less well satisfied for hot stars.

\cite{Yagi_PRD_2013} corroborated their guess about the origin of
universality by investigating the radial dependence of the integrands
entering the calculation of moment of inertia, quadrupole moment and
tidal deformability in the Newtonian limit, normalised by their
  values at the star's center, showing that they are peaked around
$0.7 \lesssim r/R \lesssim 0.9$, where $r$ is the radial distance from
the star's center. The main contribution to these quantities thus
indeed comes from the outer core and crust.  In
Fig. \ref{fig:radprofiles} we show for stars with compactness $C=0.17$
the radial profiles of the quantities
$\left( e r^4/e_c R^4 \right)$ and
  $\left[e p_c \left(de/dp \right)/e_c^2\right]$, where
$\left(e r^4 \right)$ corresponds to the integrand of the moment of
inertia and quadrupole moment in the Newtonian limit (top panel), and
$\left(e de/dr \right)$ to the EoS-dependent contribution to the tidal
deformability in the Newtonian limit (bottom panel); here index $c$
indicates corresponding values at the star's center.  Comparing
results for different $S/A$ and $Y_{L,e}$ (left) we note that with
increasing temperature and, slightly more pronounced, increasing
$Y_{L,e}$, the maximum of the integrands migrate to lower $r/R$ values
and smear out.  For the highest considered value $S/A=4$,
  $\left[e r^4/e_c R^4 \right]$ manifests a wide peak centered at
  $r/R \approx 0.6$, while
  $\left(e p_c \left(de/dp \right)/e_c^2 \right)$ shows a plateau over
  $r/R \lesssim 0.6$ followed by a shoulder-like decrease. Overall,
an important dispersion is obtained among the curves
corresponding to different thermodynamic conditions.  This is in
agreement with the breakdown of universality due to thermal effects if
different thermodynamic conditions are compared.  Considering again
fixed $S/A = 2$ and $Y_{L,e} = 0.2$ (right), the predictions of the
different models agree very well over the whole star's volume. We,
therefore, conclude that the explanation for
$I-$Love$-Q$ universality does not lie in the similar behavior of
``realistic" EoS in the outer core and the crust but rather in the
  similar behavior of EoS over the density domains which, under the
  considered thermodynamic condition, play the most important role.  

\section{Conclusions}
\label{sec:conclusions}

In this work, we constructed {EoS} of dense matter with heavy baryons
(hyperons and $\Delta$-resonances) at non-zero temperature and for
different lepton fractions {within the} covariant density functional
theory.  {This extends models of } EoS for cold $\beta$-equilibrated
matter (as it occurs in older compact stars) to finite temperatures
and matter out of $\beta$-equilibrium as needed for the description of
CCSN, PNS evolution, and BNS mergers. In particular, our
finite-temperature EoS models include heavy baryon degrees freedom -
the full baryon octet and $\Delta$-resonances.  Our EoS model is
{consistent with available} constraints from nuclear physics
experiments, {\it ab initio} calculations of low-density neutron
matter, and observations of compact stars, specifically, massive
neutron stars, radius and mass inferences by NICER experiment and
tidal deformability derived from GW170817 event.
We plan to make tables of the EoS publicly available on the Compose database.

As discussed previously for cold compact stars
\citep{Drago_PRC_2014,Li_PLB_2018}, the population of $\Delta$s at
intermediate densities, before the onset of most hyperons, leads to
smaller radii for intermediate-mass stars. The impact on the maximum
mass compared with hypernuclear models is nevertheless negligible
since at high densities anyway many different states are populated and
the additional $\Delta$ degree of freedom only leads to a
rearrangement of particle abundances.  As long as no additional
degrees of freedom are populated, {\it thermal effects lead to an
  increase of the maximum gravitational mass with $S/A$}.  This is
particularly the case of purely nucleonic stars and hypernuclear stars
at high temperatures.  The population of additional particle degrees
of freedom by thermal excitation can, on the other hand, reduce the
maximum gravitational mass with increasing $S/A$.  The lepton fraction
modifies the star's maximum gravitational mass, too.  Depending on the
EoS, $S/A$, and particle degrees of freedom, compact star's maximum
gravitational mass may increase or decrease with the lepton fraction.
Most frequently the gravitational mass of a compact star with an
admixture of heavy baryons increases with the lepton fraction, while
the opposite effect is obtained for purely nucleonic stars.  The
reason is that an increasing charge fraction decreases the charge
chemical potential and thus disfavors the appearance of, in particular
negatively charged, hyperons and $\Delta$s. Because of an extended
surface, stars at finite temperature and/or high lepton fractions are
less compact and less bound than their counterparts at zero
temperature.  The maximum baryonic mass of a hot star determines the
stability against collapse to a black hole. We have shown that,
depending on the nuclear EoS, both purely nucleonic stars, and
hypernuclear stars may be stable (unstable) within the considered
domain {of} entropy per baryon.

Several authors~\citep{Martinon_PRD_2014,Marques_PRC_2017,Lenka_JPG_2019}
have argued that thermal effects induce deviations from the universal
relations. These findings were confirmed by comparing the relations between
various global properties of compact stars at finite $S/A$ with those at
zero temperature. As a byproduct, we have shown that the $\Delta$
degrees of freedom do not alter universal relations for cold compact
stars. Finally, we have demonstrated that {\it when the universal
  relations are studied at the same entropy per baryon and the same
  lepton fraction, universality is recovered}.  We have illustrated
this by establishing universal relations between compact star's
compactness and several other global properties as well as by testing
the validity of universality for the $I-$Love$-Q$ relations.
This EoS independence could
be helpful for the analysis of observational data from hot, transient
states of compact stars, in full analogy to the zero temperature case
discussed extensively in the literature.  Our findings may also give
new hints for the understanding of the origin(s) of universality.

\section*{Acknowledgments}

This work has been partially funded by the European COST Action
CA16214 PHAROS ``The multi-messenger physics and astrophysics of
neutron stars".  A.~R.~R. acknowledges the hospitality of the
Frankfurt Institute for Advanced Studies.  A.~S.  acknowledges the
support by the Deutsche Forschungsgemeinschaft (Grant No. SE 1836/5-1)
and the hospitality of the Observatoire de Paris, Meudon.

\begin{appendix}
\section*{Appendix: Uncertainties in radii of compact stars
}
\label{sec:app}

\begin{figure}
  \begin{center}
    \includegraphics[angle=0, width=0.99\columnwidth]{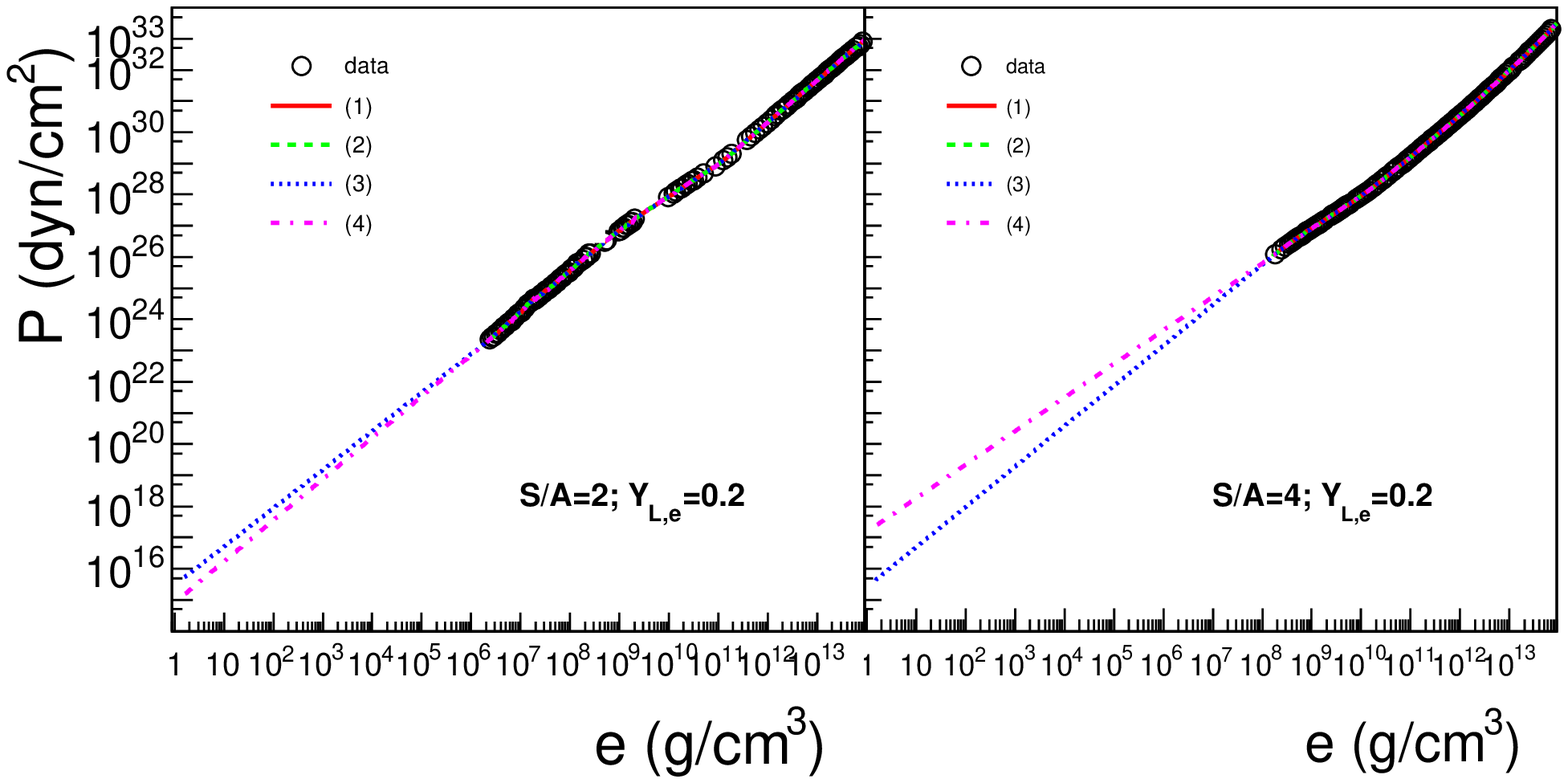}
    \includegraphics[angle=0, width=0.99\columnwidth]{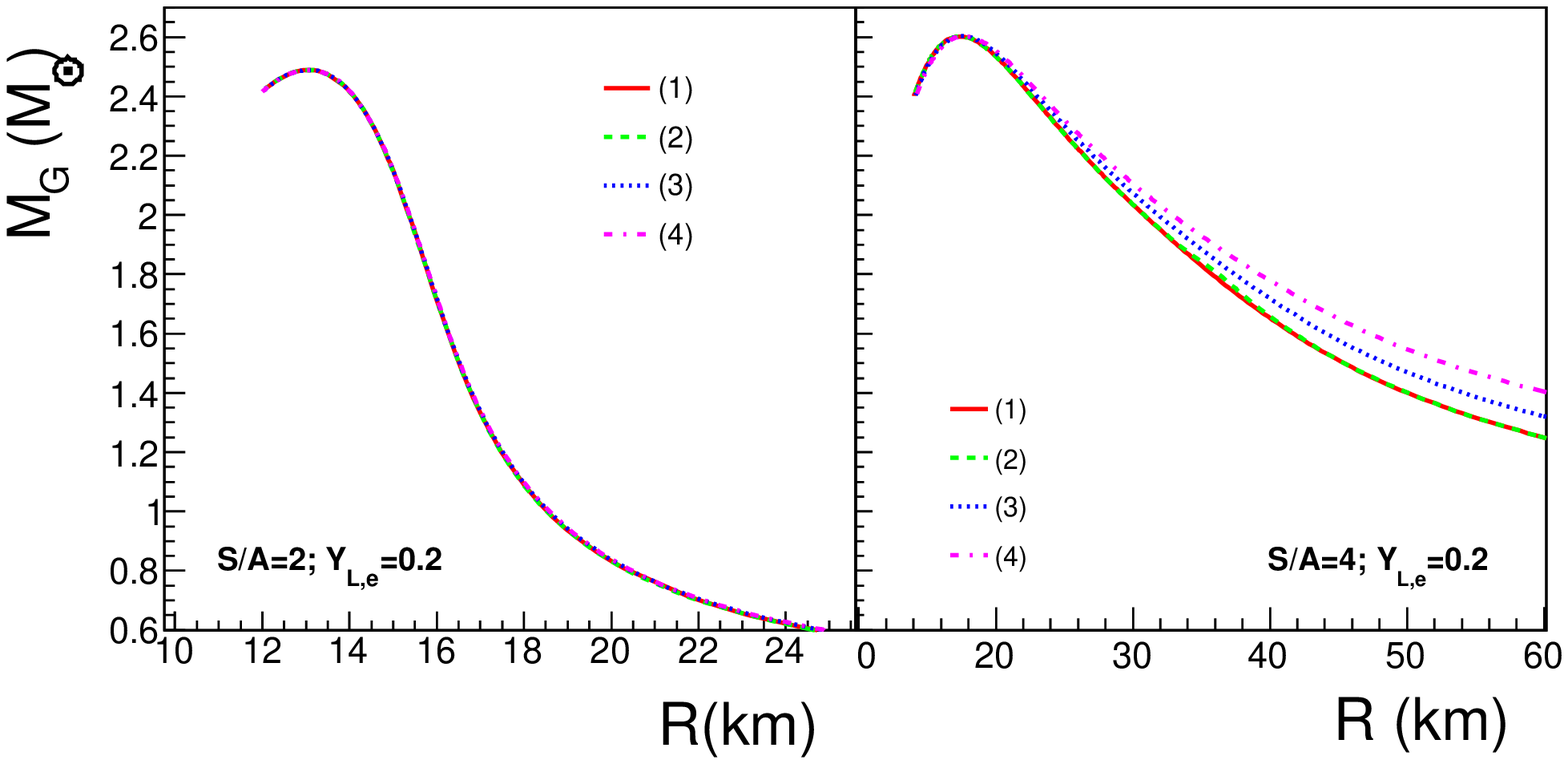}
    \includegraphics[angle=0, width=0.99\columnwidth]{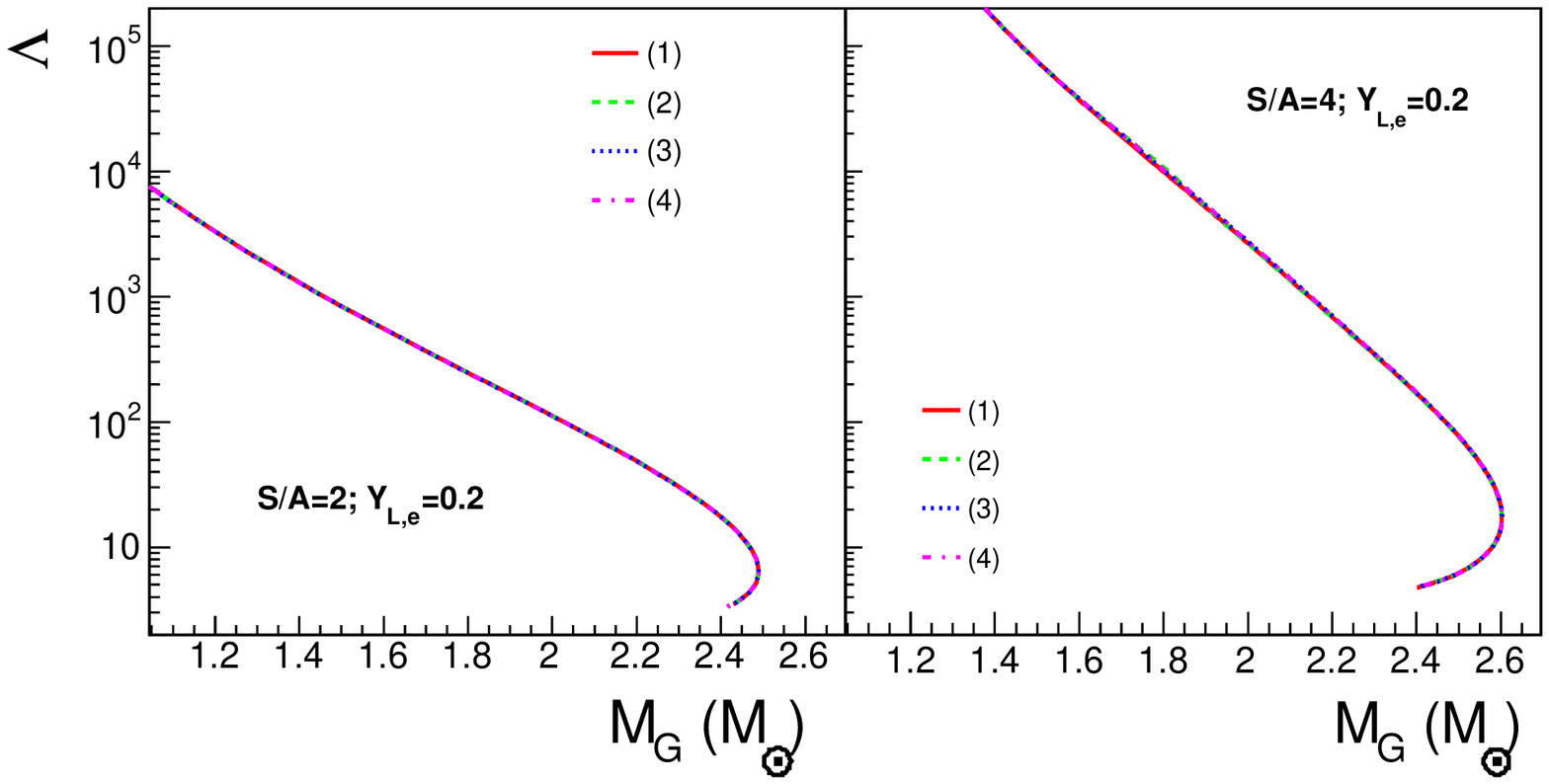}
    \includegraphics[angle=0, width=0.99\columnwidth]{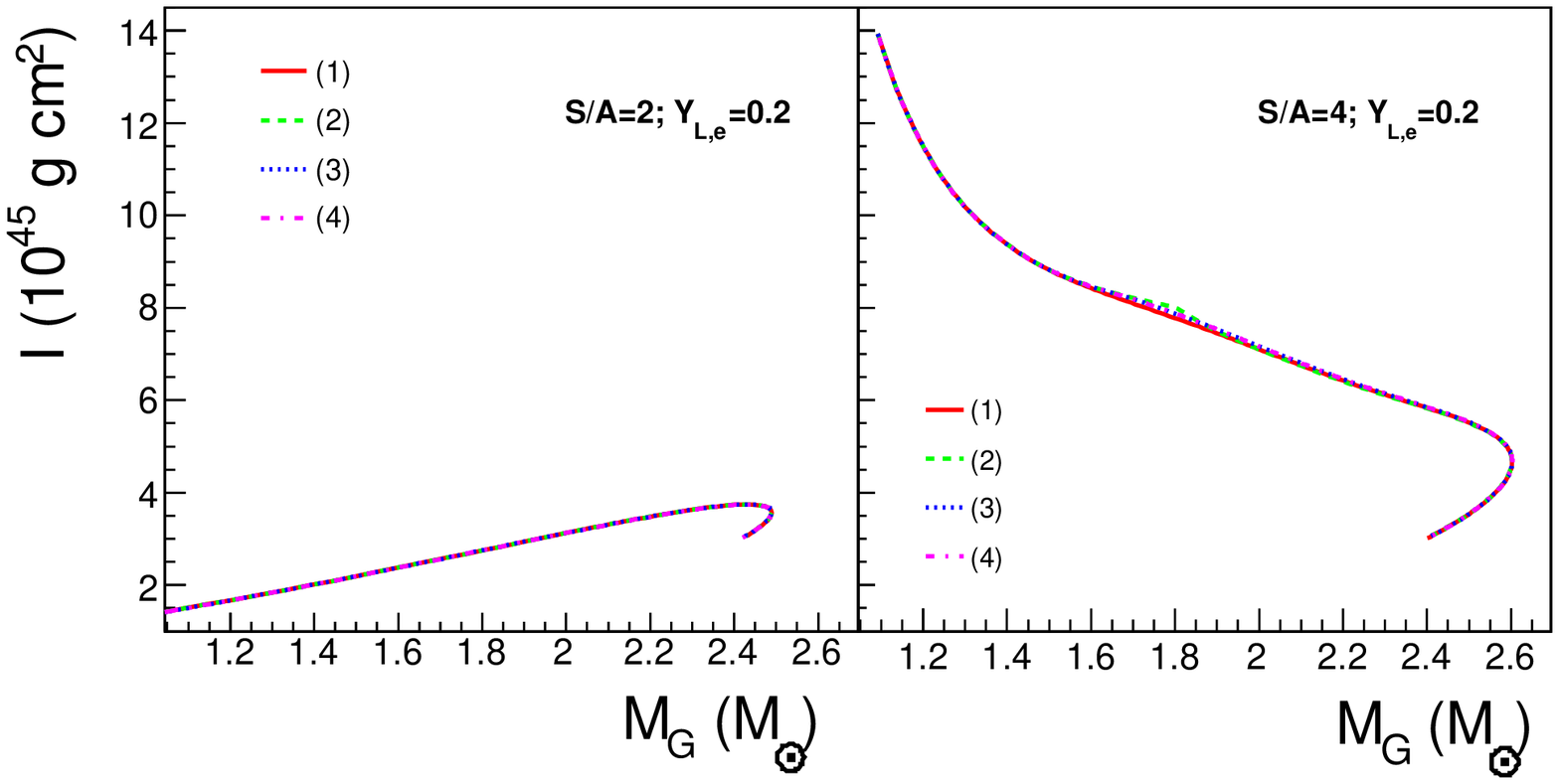}
  \end{center}
  \caption{EoS (1st row) and relations between
    gravitational mass and radius (2nd row),
    tidal deformability and gravitational mass (3rd row)
    and moment of inertia and gravitational mass (4th row)
    for non-rotating stars comparing different definitions of
    NS's surface and the crust-core transition as listed in
    Table \ref{tab:extrapol}. For the thermodynamic conditions, 
    $(S/A=2,Y_{L,e}=0.2)$ (left panels) and  $(S/A=4,Y_{L,e}=0.2)$ (right panels)
    have been chosen. The open circles marked as ``data'' indicate the EoS data
    from the Compose database corresponding
    to the DD2Y EoS~\citep{Marques_PRC_2017}.
  }
  \label{fig:uncertainties}
\end{figure}

To quantify the uncertainties related to the surface
definition and the crust-core transition we consider four different
scenarios, see Table \ref{tab:extrapol}. We thereby vary on the one
hand the (fixed) transition density from the core to the crust and on
the other hand the density at which we define the surface of the star,
implying an EoS extrapolation for cases (3) and (4).  The top panels of
Fig. \ref{fig:uncertainties} illustrate, for each scenario, the total
pressure as a function of the total energy density together with the EoS
data from the Compose database. For this example, the DD2Y
EoS~\citep{Marques_PRC_2017} has been chosen. The bottom panels depict
the mass-radius relation, where the cases  $(S/A=2,Y_{L,e}=0.2)$ (left) and
$(S/A=4,Y_{L,e}=0.2)$ (right) have been considered.

As indicated in the main text, we find that: i) the uncertainties related
to the arbitrarily chosen value of the crust-core transition density
$n_t$ are negligible, ii) the uncertainties related to the surface
definition are sizeable only for stars with high values of $S/A
\gtrsim 3$. For the shown example, at $S/A = 4, Y_{L,e} = 0.2$, the
uncertainty in the radius of a 1.4 $M_\odot$ star amounts to 20\%.
This indicates an upper limit on the uncertainty; for all other
studied star properties, the uncertainty is smaller. Similar
uncertainties are found for other EoS models, showing that our results
are robust.

\begin{table}
  \begin{tabular}{cccl}
    \hline 
    case & $n_{min}$   & $n_{t}$ & $n_B$ fit domain \\
    \hline
    (1)  & $n_{ll}$    &   $n_s/3$   & - \\
    (2)  & $n_{ll}$    &   $2 n_s/3$   & - \\
    (3)  & $10^{-15}$ [fm$^{-3}$] &   $n_s/2$   & $(n_{ll},n_{t})$ \\
    (4)  & $10^{-15}$ [fm$^{-3}$] &   $n_s/2$   & $(n_{ll},n_{LIN})$ \\
 \hline
  \end{tabular}
  \caption{Definition of NS surface and matching between crust and
    core.  $n_{min}$ defines the baryon number density at the star's
    surface, $n_{t}$ the transition density from core to crust,
    $n_{ll}$ the lowest density in the EoS data tables for the given
    conditions and $n_{\mathrm{LIN}}$ the maximum density for which
    $\log(e)-\log(n_B)$ and $\log(P)-\log(n_B)$ show a linear beahvior
    to a very good precision between $n_{ll}$ and $n_{\mathrm{LIN}}$;
    $n_s$ stands for the saturation density of symmetric nuclear
    matter.  }
  \label{tab:extrapol}
\end{table}

\end{appendix}

\bibliographystyle{mnras}
\bibliography{NYD.bib}
\label{lastpage}
\end{document}